\documentclass[english,aps,manuscript,preprintnumbers,superscriptaddress,nofootinbib]{revtex4}
\usepackage[T1]{fontenc}
\usepackage[latin9]{inputenc}
\usepackage{amsmath}
\usepackage{amssymb}
\usepackage{wasysym}
\usepackage{graphicx}
\usepackage{esint}

\usepackage[marginal]{footmisc}

\renewcommand{\baselinestretch}{1.1}

\makeatletter

\@ifundefined{textcolor}{}
{%
 \definecolor{BLACK}{gray}{0}
 \definecolor{WHITE}{gray}{1}
 \definecolor{RED}{rgb}{1,0,0}
 \definecolor{GREEN}{rgb}{0,1,0}
 \definecolor{BLUE}{rgb}{0,0,1}
 \definecolor{CYAN}{cmyk}{1,0,0,0}
 \definecolor{MAGENTA}{cmyk}{0,1,0,0}
 \definecolor{YELLOW}{cmyk}{0,0,1,0}
}

\makeatother

\usepackage{babel}
\begin{document}

\preprint{}

\title{Long-range Self-interacting Dark Matter in the Sun}

\author{Jing Chen}
\email{jchen@itp.ac.cn}
\affiliation{University of Chinese Academy of Science \\19A Yuquan Road, Beijing, 100049, P.R. China}
\affiliation{State Key Laboratory of Theoretical Physics,\\Kavli Institute for Theoretical Physics China,\\Institute of Theoretical Physics, Chinese Academy of Science\\Zhong Guan Cun Street 55\#, Beijing, 100190, P.R. China}

\author{Zheng-Liang Liang}
\email{liangzl@itp.ac.cn}
\affiliation{Institute of High Energy Physics, Chinese Academy
of Science\\19B Yuquan Road, Beijing, 100049, P.R. China}

\author{Yue-Liang Wu}
\email{ylwu@itp.ac.cn}
\affiliation{State Key Laboratory of Theoretical Physics,\\Kavli Institute for Theoretical Physics China,\\Institute of Theoretical Physics, Chinese Academy of Science\\Zhong Guan Cun Street 55\#, Beijing, 100190, P.R. China}

\author{Yu-Feng Zhou}
\email{yfzhou@itp.ac.cn}
\affiliation{State Key Laboratory of Theoretical Physics,\\Kavli Institute for Theoretical Physics China,\\Institute of Theoretical Physics, Chinese Academy of Science\\Zhong Guan Cun Street 55\#, Beijing, 100190, P.R. China}

\begin{abstract}
We investigate the implications of the long-rang self-interaction on both
the self-capture and the annihilation of the
self-interacting dark matter (SIDM) trapped in the Sun. Our discussion is   based on a specific SIDM model in which DM particles self-interact via a light scalar mediator, or Yukawa potential, in the context of quantum mechanics. Within this framework, we calculate the self-capture rate   across a broad  region of parameter space. While the self-capture rate can be obtained  separately in the Born regime with perturbative method, and in the classical limits with the Rutherford formula, our calculation covers the gap between in a non-perturbative fashion. Besides, the phenomenology of  both the Sommerfeld-enhanced $s$- and
$p$-wave annihilation of  the solar SIDM is also involved in our discussion.  Moreover, by combining
the analysis of the Super-Kamiokande (SK) data and the observed DM relic density,
we constrain the nuclear capture rate of the DM particles in the presence of the dark Yukawa potential. The consequence of  the  long-range dark force on probing the solar SIDM turns out to be significant if the force-carrier is much lighter than the DM particle, and a quantitative analysis is provided.
\end{abstract}
\maketitle

\section{Introduction\label{sec:Section-I}}

Although the existence of dark matter (DM) have been well established by
the observations from the galactic scale up to the cosmological scale,
the particle nature of DM still remains unclear.
While the collisionless Weakly Interacting Massive Particle (WIMP) has been
the focus of the current DM study,
the possibility of the self-interacting dark matter (SIDM) has also
drawn increasing attention in both  particle physics and astrophysics
(see e.g.
\citep{1992ApJ...398...43C,1995ApJ...452..495D,Hogan:2000bv,2001ApJ...547..574D,2004PhRvL..92c1303H,2005PhRvD..71f3528H}).
Long-range self-interactions between DM particles induced by
the exchange of  light mediator particles can lead to
the Sommerfeld enhancement of DM annihilation cross sections
\cite{Sommerfeld:1931,
Hisano:2002fk,
Hisano:2003ec,
ArkaniHamed:2008qn,
Feng:2009hw,Feng:2010zp%
},
which can be used to explain the observed cosmic-ray positron excesses
\cite{
Adriani:2008zr,
Adriani:2010ib,
FermiLAT:2011ab,
Accardo:2014lma%
}
by providing a source of the boost factor of $\mathcal{O}(10^{2}-10^{3})$
(for a recent global analysis on the boost factor, see e.g.
\cite{
Kopp:2013eka,
Yuan:2013eja,
Cholis:2013psa,
Jin:2013nta,
Liu:2013vha,
Ibarra:2013zia,
Jin:2014ica}).
It is under active investigation whether an excess of antiproton also exists
in the AMS-02 $\bar p/p$ data
\cite{Giesen:2015ufa,
Jin:2015sqa,
Kohri:2015mga,
Chen:2015kla,
Chen:2015cqa,
Lin:2015taa,
Hamaguchi:2015wga,
Ibe:2015tma}
and a boost factor is also required in the antiproton channel.
Phenomenological consequences of the light mediator exchange between DM particles in
astrophysics and cosmology have been
discussed in a series of studies~\cite{Lattanzi:2008qa,ArkaniHamed:2008qn,
Feng:2009hw,Feng:2009mn,Slatyer:2009vg,Buckley:2009in,Dent:2009bv,Zavala:2009mi,
Feng:2010zp,Hannestad:2010zt,Hisano:2011dc,Hooper:2012cw,
Bellazzini:2013foa,Tulin:2013teo,Laha:2013gva,Gabrielli:2013jka,Foot:2014osa,Foot:2014uba,
Lopes:2014aoa,Laha:2015yoa,Blennow:2015xha} 
which include its implications on
relic density~\cite{
Dent:2009bv,
Zavala:2009mi,
Feng:2009hw,
Feng:2010zp},
CMB~\cite{
Zavala:2009mi,
Hannestad:2010zt,
Hisano:2011dc},
halo shape~\cite{Tulin:2013teo,Fan:2013yva,
Buckley:2009in},
and
small scale structures~\cite{
Lattanzi:2008qa,
Robertson:2009bh,Foot:2014uba},
etc.

Another interesting phenomenology associated with the DM self-interaction
is its self-capture within the Sun. In addition to solar elements,
DM particles that have already been captured within the Sun may also contribute
to the capture of the passing halo ones. Consequently, more DM particles will build up within the Sun and an enhancement of the high-energy neutrino signals from the DM annihilation might be possible. Under the assumption of a constant self-scattering cross section, the authors of Ref.~\citep{2009PhRvD..80f3501Z,Albuquerque:2013xna,Chen:2014oaa}
 utilize neutrino detectors to probe the solar SIDM particles
 and explore for the viable parameter space. It is found that  a significant enhancement of the solar neutrino flux requires a disparity between  two separate input parameters, $i.e.$, a large DM self-capture rate and a small annihilation cross section. In this paper, we will
extend this case to the scenario where the DM particles interact with each other
via a light mediator (also in Ref.~\citep{Fan:2013bea,Chen:2015bwa}\footnote[1]{\renewcommand{\baselinestretch}{1}\selectfont In these two papers, the self-capture rate is calculated either in the Coulomb limit~\citep{Fan:2013bea} or with the perturbative method~\citep{Chen:2015bwa}.}). In such scenario, this force-carrier
 will play a role in both the self-capture and annihilation
within the Sun, as well as in thermal freeze-out that accounts for
the observed DM density. As a result, the self-capture and annihilation rate become inter-related to each other, and more than that, the disparity between them may be achieved in a natural way. To investigate the relevant implications from
an optimistic view, we examine a concrete SIDM model in which the Majorana
DM particles $\chi$ interact with each other through the exchange of  a light scalar mediator $\phi$, so that
the ``irreducible'' annihilation process $\chi\chi\rightarrow\phi\phi$
proceeds through the $p$-wave channel. The $p$-wave annihilation is of particular interest because contrary
to its $s$-wave counterpart that is subject to the Sommerfeld
enhancement, it suffers a velocity-suppression not only in the Sun but at the kinetic decoupling stage, and hence results in a larger relic density~\citep{Feng:2010zp}.
As a consequence, a larger self-interaction coupling $\alpha_{\chi}$ is
required accordingly for the $p$-wave annihilation to give the correct relic
density, which is also a prerequisite for a large self-capture rate.

Following the literature~\citep{Feng:2009hw,Buckley:2009in,Tulin:2013teo}
in which authors managed to discuss the DM self-scattering beyond the perturbative
level, we similarly give a quantitative description of the self-capture
rate for the Majorana DM particles. However, instead of adopting the transfer cross
section as a proxy quantity to describe the self-scattering, we calculate
the self-capture rate with long-range interaction in a straightforward
manner, considering that the kinetic requirement for self-capture
automatically regulates the forward-scattering divergence and hence
makes the self-capture rate a well-defined quantity.

In addition, in order to explore the phenomenology of the solar SIDM from a comprehensive
perspective, our discussion  includes  the Sommerfeld-enhanced
annihilation of the solar SIDM as well.  We illustrate that while the DM particles predominantly annihilate
to light mediators through the $p$-wave channel, a small but non-vanishing
quota of annihilation proceeding through the $s$-wave channel in
the thermal freeze-out may become dominant over its $p$-wave counterpart
within the Sun, and produce signals at the terrestrial neutrino telescopes.
For such case we use the neutrino telescope Super-Kamiokande~\citep{Choi:2015ara}
to find out whether  the Sommerfeld effect can help constrain the nuclear capture rate $C_{\mathrm{n}}$
in a nontrivial manner. These constraints can be translated into upper limits
on the DM-nucleon cross section for other SIDM models. It is found that in the approach of the Coulomb limit of the long-range self-interaction, the constraints from the Super-Kamiokande on nuclear capture rate can be significantly tightened.

This paper is organized as follows. In Sec.~\ref{sec:Model-Setup}
we take an introduction to the model setup for our study. In Sec.~\ref{sec:2} we introduce
the related background about the solar DM, including its self-capture
and evolution within the Sun. We discuss the implications of the light mediator on the self-capture and annihilation of the solar DM particles, as well
as on relic density in Sec~\ref{sec:3}. We further investigate its
impact on the neutrino telescope Super-Kamiokande in Sec.~\ref{sec:4}, and
conclusions are drawn in Sec.~\ref{sec:5}.

\section{Model Setup\label{sec:Model-Setup}}
We are interested in the the dark sector that gives rise to the Yukawa
potential of the DM self-interaction. As a familiar
prototype, Yukawa potential between DM particles provides us with a representative
description and a concrete example of the long-range self-interaction~\citep{ArkaniHamed:2008qn,Feng:2010zp,Tulin:2013teo}. The self-interacting
Yukawa potential\footnote[1]{\renewcommand{\baselinestretch}{1}\selectfont Strictly speaking, Yukawa potential narrowly refers to the attractive interaction in  Eq.~(\ref{eq:Yukawa}). However, for the purpose of convenience, here we  use it to refer to  both the attractive and repulsive interactions.} is written as
\begin{equation}
V\left(r\right)=\pm\frac{\alpha_{\chi}}{r}e^{-m_{\phi}r},\label{eq:Yukawa}
\end{equation}
which describes in non-relativistic limit the following interactions
of the dark sector \citep{Tulin:2013teo}:
\begin{eqnarray}
\mathcal{L}_{\chi\phi} & = & \begin{cases}
g_{\chi}\bar{\chi}\gamma^{\mu}\chi\phi_{\mu} & \mathrm{vector}\,\mathrm{mediator}\\
g_{\chi}\bar{\chi}\chi\phi & \mathrm{scalar}\,\mathrm{mediator}
\end{cases},\label{eq:interaction}
\end{eqnarray}
where $m_{\phi}$ is the mass of the mediator particle $\phi$, and
the dark fine-structure constant $\alpha_{\chi}$ is connected to the coupling
$g_{\chi}$ through $\alpha_{\chi}=g_{\chi}^{2}/\left(4\pi\right)$.
For a vector force-carrier, the interaction between DM could be attractive
($\bar{\chi}\chi$ scattering) or repulsive ($\chi\chi$ or $\bar{\chi}\bar{\chi}$
scattering), while the force can only be attractive ($-$) between
DM for the scalar mediator scenario.
\begin{figure}
\begin{centering}
\includegraphics[scale=0.6]{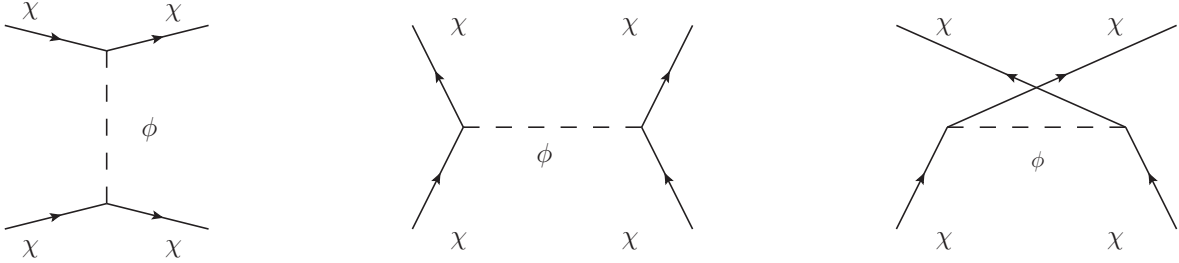}
\par\end{centering}

\protect\caption{\label{fig:The-tree-level-diagrams}The tree-level diagrams for elastic
scattering process $\chi\chi\rightarrow\chi\chi$}
\end{figure}
Since in this paper we are interested in spin-$1/2$ identical DM,
i.e., Majorana DM particles, the interaction between DM and the vector
mediator in Eq.~(\ref{eq:interaction}) is absent, and an extra $s$-channel
process should be included in the calculation of relevant scattering
amplitude for a scalar mediator $\phi$, which is shown in Fig.~\ref{fig:The-tree-level-diagrams}.
While the second and the third diagrams in Fig.~\ref{fig:The-tree-level-diagrams}
correspond to the long-range Yukawa interaction in non-relativistic
limit, the first $s$-channel process can be described non-relativistically
with a contact interaction, which is subject to the Sommerfeld enhancement
that will be reviewed in detail below. However, due to its highly
suppressed cross section $\sigma_{\chi\chi}\sim\alpha_{\chi}^{2}v_{\mathrm{rel}}^{3}/m_{\chi}^{2}$,
contribution of the $s$-channel process can be safely ignored compared
with that of the Yukawa potential, even if the Sommerfeld enhancement
is involved. So only the Yukawa potential is relevant for the DM scattering.

Although we want to explore phenomenologically the impact of the DM
long-range self-interaction as far as possible, it is unrealistic for our discussion to be entirely
model-independent.  At the very least, the phenomenological consequence
of the irreducible annihilation process $\chi\chi\rightarrow\phi\phi$
has to be taken into consideration (see Fig.~\ref{fig:annihilation}). So, we study a model that not only gives
rise to significant Sommerfeld effects within the Sun, but also cover
as wide a range of phenomenology as possible, namely, the self-capture
and both the $s$- and $p$-wave annihilation of the solar SIDM.

To facilitate the novelty brought by the long range
self-interaction, we assume that the irreducible $p$-wave process
(shown in Fig.~\ref{fig:annihilation}) dominates the annihilation
during the thermal freeze-out. The reason for this assumption is two-fold:
$\left.1\right)$ As opposed to the $s$-wave-dominated case, the
$p$-wave-dominated annihilation avoids severe constraint on the coupling
$\alpha_{\chi}$ imposed by the DM relic density, so that the Sommerfeld
effect remains significant~\citep{Feng:2010zp,Chen:2013bi}. $\left.2\right)$
The $p$-wave-dominated annihilation within the Sun is subject to
such significant velocity-suppression that the total annihilation
rate $\Gamma_{A}$ will be enhanced remarkably (see Eq.~(\ref{eq:annihilation rate})
and~(\ref{eq:Annihilation1}) in the following section), which may further favor the indirect
detection of the solar DM.

On the other hand, given the dark sector of the Majorana DM $\chi$ and the scalar mediator
$\phi$, it is natural to assume that they couple to the SM sector
through the Higgs portal~\citep{Burgess:2000yq,Patt:2006fw,Andreas:2008xy,Andreas:2010dz,Djouadi:2011aa,Pospelov:2011yp,Greljo:2013wja}. In addition, extra four-fermion interactions
are introduced to account for the  DM $s$-wave annihilation into the SM
particles.
Integrating all these considerations, we consider the following interaction as an example:
\begin{figure}
\begin{centering}
\includegraphics[scale=0.6]{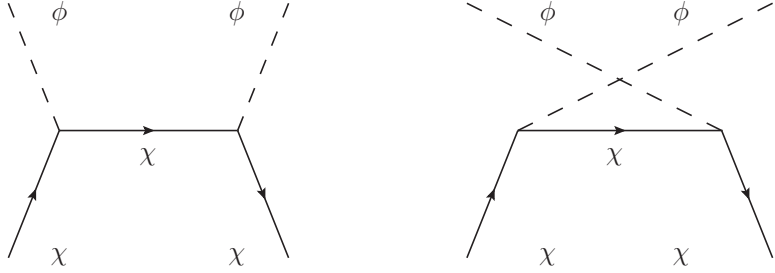}
\par\end{centering}

\protect\caption{\label{fig:annihilation} The diagrams for the minimal annihilation
process $\chi\chi\rightarrow\phi\phi$. }
\end{figure}
\begin{eqnarray}
\mathcal{L} & \supset & \frac{g_{\chi}}{2}\bar{\chi}\chi\phi-\sum_{f}y_{f}\sin\theta\,\phi\bar{f}f+y_{N}\phi N^{c}N+\sum_{f}\frac{G_{f}}{2}i\bar{\chi}\gamma^{5}\chi\bar{f}f,\label{eq:model Interacton}
\end{eqnarray}
where $\sin\theta$ describes the mixing between $\phi$ and Higgs boson,
and $y_{f}=m_{f}/v_{\mathrm{EW}}$ is the Yukawa coupling of the SM fermion $f$,
with the SM fermion mass $m_{f}$ and the Higgs vacuum expectation
value $v_{\mathrm{EW}}\approx246\,\mathrm{GeV}$. $G_{f}$ is the
coupling strength between the DM and the SM fermion $f$. The coupling
of $\phi$ and right-handed sterile neutrino $N$ is introduced in
Ref.~\citep{Kouvaris:2014uoa} to save the light mediator $\phi$
from the Big Bang Nucleosynthesis (BBN) constraints. To avoid the
overclosure problem, overproduction of entropy and changes to the
light element abundances during the BBN, $\phi$ is required to decay
before the BBN ($t\sim1\,\mathrm{s}$)~\citep{Kaplinghat:2013yxa},
which corresponds to the bound~\citep{Kouvaris:2014uoa}
\begin{equation}
y_{N}\apprge6\times10^{-12}\left(\frac{100\,\mathrm{MeV}}{m_{\phi}}\right)^{1/2}.
\end{equation}

\section{Dark matter in the Sun\label{sec:2}}

\subsection{\label{sub:High-energy-neutrinos}Dark matter in the Sun}

The accumulation of the DM in the sun is initiated by DM-nuclei collisions
that cost the passing-by Galactic DM particles sufficient kinetic energy to
escape from the gravitational pull of the Sun. In this process, subsequent
collisions tend to further lower the kinetic energy of the DM rather
than eject them back into the deep space as long as the DM mass is
heavier than a few GeV, hence the number of the trapped DM grows until
their capture and annihilation  reach an equilibrium. If the DM self-interaction
is taken into account, the evolution of the solar DM number $N_{\chi}$
is depicted by the following equation
\begin{equation}
\frac{dN_{\chi}}{dt}=C_{\mathrm{n}}+\left(C_{\mathrm{s}}-E_{\mathrm{n}}\right)N_{\chi}-\left(C_{\mathrm{a}}+E_{\mathrm{s}}\right)N_{\chi}^{2},\label{eq:DMnumber equation}
\end{equation}
where $C_{\mathrm{n}}$ is the DM capture rate by (scattering off) nuclei in the Sun, $C_{\mathrm{a}}$ is twice the annihilation rate
of a pair of DM particles, $E_{\mathrm{n}}$ and $E_{\mathrm{s}}$ represent respectively the evaporation
rate by nuclei and by the captured DM particles, and $C_{\mathrm{s}}$ is the capture rate
by the trapped DM particles, a key physical quantity of our concern in this paper.
It should be noted that in the range of the DM mass of our interest
$\left(m_{\chi}\geq10\,\mathrm{GeV}\right)$ it is justified for us
to omit the the evaporation rate $E_{\mathrm{n}}$ in Eq.~(\ref{eq:DMnumber equation}), since it  is irrelevant for a DM mass
larger than $5\,\mathrm{GeV}$~\citep{Griest:1986yu,Gould:1987ir,Busoni:2013kaa}.\footnote[1]{\renewcommand{\baselinestretch}{1}\selectfont For smaller DM masses, the detection of the solar DM evaporation is discussed in Ref.~\citep{Kouvaris:2015nsa}.}
Besides, we also omit the the self-evaporation rate $E_{\mathrm{s}}$ because $C_{\mathrm{a}}\gg E_{\mathrm{s}}$,
which will be demonstrated in detail in Appendix \ref{sec:appendixA}.
Thus Eq.~(\ref{eq:DMnumber equation}) is reduced to
\begin{equation}
\frac{dN_{\chi}}{dt}=C_{\mathrm{n}}+C_{\mathrm{s}}N_{\chi}-C_{\mathrm{a}}N_{\chi}^{2},
\end{equation}
which has an analytic solution

\begin{equation}
N_{\chi}=\frac{C_{\mathrm{n}}\tanh\left(t\cdot\xi\right)}{\xi-\left(C_{\mathrm{s}}/2\right)\tanh\left(t\cdot\xi\right)},\label{eq:DMnumber}
\end{equation}
with

\begin{equation}
\xi=\sqrt{C_{\mathrm{n}}\cdot C_{\mathrm{a}}+C_{\mathrm{s}}^{2}/4}.
\end{equation}
If $\tanh\left(t_{\mathrm{\odot}}\cdot\xi\right)\simeq1$ in Eq.~(\ref{eq:DMnumber})
with solar age $t_{\mathrm{\odot}}=4.5\times10^{9}\,\mathrm{yr}$
, the capture and annihilation of the DM reach an equilibrium today.
 Then the annihilation rate $\varGamma_{A}$ for a steady number of
DM within the Sun can be written as
\begin{eqnarray}
\varGamma_{A} & = & \frac{1}{2}C_{\mathrm{a}}N_{\chi}^{2}\nonumber \\
 & = & \frac{1}{2}C_{\mathrm{a}}\left(\sqrt{\frac{C_{\mathrm{n}}}{C_{\mathrm{a}}}+\frac{C_{\mathrm{s}}^{2}}{4C_{\mathrm{a}}^{2}}}+\frac{C_{\mathrm{s}}}{2C_{\mathrm{a}}}\right)^{2}\nonumber \\
 & = & \frac{1}{2}\left(C_{\mathrm{n}}+\frac{C_{\mathrm{s}}^{2}}{2C_{\mathrm{a}}}+\sqrt{\frac{C_{\mathrm{n}}}{C_{\mathrm{a}}}+\frac{C_{\mathrm{s}}^{2}}{4C_{\mathrm{a}}^{2}}}\cdot C_{\mathrm{s}}\right),\label{eq:annihilation rate}
\end{eqnarray}
which is reduced to the trivial case where the dark matter self-interaction
is absent, i.e., when $\varGamma_{A}=C_{\mathrm{n}}/2$ if $C_{\mathrm{s}}^{2}/(C_{\mathrm{n}}C_{\mathrm{a}})\ll1$
is satisfied.

\subsection{Self-capture and annihilation of the self-interacting dark matter}

The formula for the self-capture rate can be obtained after a minor
modification to the procedures in Ref.~\citep{Gould:1987ir,Gould:1991hx}
as the following:
\begin{equation}
C_{\mathrm{s}}N_{\chi}=\frac{\rho_{\chi}}{m_{\chi}}\int_{\mathrm{Sun}}n_{\chi}\left(r\right)d^{3}r\int\sigma_{\mathrm{sc}}\left(w\right)\frac{w^{2}}{u}f\left(\mathbf{u}\right)d^{3}u,\label{eq:Cs1}
\end{equation}
where $\rho_{\chi}=0.3\,\mathrm{GeV}$ is the DM local density in
our solar neighborhood, $m_{\chi}$ manifestly the mass of the DM.
$f\left(\mathbf{u}\right)$ is the velocity distribution of DM in
the Sun's rest frame, which is expressed as
\begin{equation}
f\left(\mathbf{u}\right)=\frac{e^{-\frac{\left(\mathbf{u}+\mathbf{v}_{\mathrm{\odot}}\right)^{2}}{v_{0}^{2}}}}{N(v_{\mathrm{esc}})},
\end{equation}
where $\left|\mathbf{v}_{\mathrm{\odot}}\right|=220\,\mathrm{km\cdot s^{-1}}$
is the velocity of the Sun, $\mathbf{u}$ being the DM velocity at infinity where the
solar gravitational effects is negligible, $v_{0}=220\,\mathrm{km\cdot s^{-1}}$
the DM velocity dispersion, and $N(v_{\mathrm{esc}})$ is the normalization
constant dependent on the Galactic escape velocity $v_{\mathrm{esc}}=544\,\mathrm{km\cdot s^{-1}}$. $w=\sqrt{v_{\mathrm{esc}}^{2}\left(r\right)+u^{2}}$ is the velocity of the incident DM particle, with $v_{\mathrm{esc}}\left(r\right)$
the solar escape velocity at radius $r$.
The DM distribution within the Sun can be described with a characteristic
temperature $T_{\chi}$ as follows\footnote[1]{\renewcommand{\baselinestretch}{1}\selectfont To avoid distraction we postpond a discussion on the solar SIDM distribution to Appendix~\ref{sec:appendixDMDistribution}.}
\begin{equation}
n_{\chi}\left(r\right)\varpropto\exp\left[-\frac{m_{\chi}}{T_{\chi}}\int_{0}^{r}\frac{GM\left(r'\right)}{\, r'^{2}}dr'\right],
\end{equation}
where $G$ is the Newton's constant and $M\left(r\right)$ represents
the solar mass contained within radius $r$. $T_{\chi}$ is used to
be determined self-consistently through the following equation:
\begin{equation}
T_{\chi}=T_{\mathrm{\odot}}(R_{\chi}),\label{eq:EffectiveTemperature}
\end{equation}
in which the right hand side represents the solar temperature at the
mean value of DM radius
\begin{equation}
R_{\chi}=\frac{\int_{\mathrm{Sun}}n_{\chi}\left(r\right)rd^{3}r}{\int_{\mathrm{Sun}}n_{\chi}\left(r\right)d^{3}r}.
\end{equation}
We use the Standard Sun Model (SSM) GS98~\citep{Serenelli:2009yc}
to obtain a fitting function of $R_{\chi}$ as follows
\begin{equation}
R_{\chi}\simeq0.012\times\sqrt{\frac{100\,\mathrm{GeV}}{m_{\chi}}}R_{\mathrm{\odot}},\label{eq:DMradius}
\end{equation}
with solar radius $R_{\mathrm{\odot}}$. The effective cross section
for self-capture, $\sigma_{\mathrm{sc}}\left(w\right)$, is dependent
on the  velocity of the incoming DM particle $w$,
if we approximate the target DM trapped within the Sun to be at rest
compared with the accelerated Galactic ones. On the other hand, for the mass range concerned in this
paper ($m_{\chi}\geq10\,\mathrm{GeV}$), we further assume that the
captured DM particles are located at the center of the Sun (see Eq.~(\ref{eq:DMradius})), thus the self-capture
rate can be simplified as
\begin{equation}
C_{\mathrm{s}}=\frac{\rho_{\chi}}{m_{\chi}}\int\sigma_{\mathrm{sc}}\left(w\right)\frac{w^{2}}{u}f\left(\mathbf{u}\right)d^{3}u.\label{eq:CsFormula}
\end{equation}
where $w=\sqrt{v_{\mathrm{c}}^{2}+u^{2}}\equiv\sqrt{v_{\mathrm{esc}}^{2}\left(0\right)+u^{2}}$.

However, here we note that instead of growing infinitely with the
accretion of the captured DM particles,  $C_{\mathrm{s}}N_{\chi}$
saturates to a maximum value set by the geometric limit of the resident DM, above which  $C_{\mathrm{s}}N_{\chi}$ is replaced by
\begin{eqnarray}
\left(C_{\mathrm{s}}N_{\chi}\right)_{\mathrm{max}} & = & \frac{\rho_{\chi}}{m_{\chi}}\pi R_{\chi}^{2}\int\left(\frac{v_{\mathrm{esc}}^{2}\left(R_{\chi}\right)+u^{2}}{u}\right)f\left(\mathbf{u}\right)d^{3}u\nonumber \\
 & \simeq & \frac{\rho_{\chi}}{m_{\chi}}\pi R_{\chi}^{2}\int\left(\frac{v_{\mathrm{c}}^{2}+u^{2}}{u}\right)f\left(\mathbf{u}\right)d^{3}u\nonumber \\
 & \simeq & 4.67\times10^{24}\times\left(\frac{100\,\mathrm{GeV}}{m_{\chi}}\right)^{2}\cdot\mathrm{s}^{-1}.
\end{eqnarray}
On the other hand, the annihilation rate of the solar DM can be expressed as
\begin{eqnarray}
C_{\mathrm{a}} & = & \left\langle \sigma_{\mathrm{ann}}v_{\mathrm{rel}}\right\rangle _{\odot}\frac{\int_{\mathrm{Sun}}n_{\chi}^{2}\left(r\right)d^{3}r}{\left(\int_{\mathrm{Sun}}n_{\chi}\left(r\right)d^{3}r\right)^{2}},\nonumber \\
  & \equiv & \frac{\left\langle \sigma_{\mathrm{ann}}v_{\mathrm{rel}}\right\rangle _{\odot}}{V_{\mathrm{eff}}},\label{eq:Annihilation1}
\end{eqnarray}
where $\left\langle \sigma_{\mathrm{ann}}v_{\mathrm{rel}}\right\rangle _{\odot}$
is the thermal average of the DM annihilation cross section times
relative velocity $v_{\mathrm{rel}}$ within the Sun, and the effective
volume occupied by the trapped DM particles can also be described with a fitting
function as follows:
\begin{equation}
V_{\mathrm{eff}}=6.9\times10^{27}\left(\frac{100\,\mathrm{GeV}}{m_{\chi}}\right)^{3/2}\,\mathrm{cm^{3}}.
\end{equation}

\section{Implications of the long-range self-interacting DM\label{sec:3}}

In this section, we study the implications of the long-range self-interaction
on the self-capture and annihilation within the Sun. Due to the multiple
exchanges of the light force-carrier between DM particles, the relevant
calculations have to be carried out beyond the perturbative approach.
The quantum mechanical description of the two-body scattering is proved
to be an effective framework to do so, which is connected to the self-capture
and annihilation in a manner that the asymptotic wave-function at infinity
describes the final state of the scattering, and the wave-function
at origin partly determines the probability of the annihilation. Now we
delve into the details.

\subsection{Calculation of the self-capture rete \label{sub:subsectionA}}

We calculate the self-capture cross section $\sigma_{\mathrm{sc}}\left(w\right)$
in Eq.~(\ref{eq:Cs1}) in the center-of-mass (CM) frame, in which
the self-capture differential cross section for a pair of identical
DM can be expressed as (after averaging over the initial spins and
summing over the final spins )
\begin{equation}
\frac{d\sigma_{\mathrm{sc}}}{d\Omega}=\frac{1}{2}\left[\sum_{S}\left(2S+1\right)\right]^{-1}\cdot\left(\sum_{S}\left|f\left(\theta\right)+\left(-1\right)^{S}\cdot f\left(\pi-\theta\right)\right|^{2}\right),\label{eq:differentialCS}
\end{equation}
where $S$ represents the total spin of the DM-pair, the factor $1/2$
accounting for the symmetry of the out-going DM particles, and the elastic scattering
amplitude $f\left(\theta\right)$ is related to the wave function
in position space via
\begin{eqnarray}
\left\langle \mathbf{x}\left|\psi_{\mathbf{p}}^{+}\right.\right\rangle  & = & \frac{1}{\left(2\pi\right)^{3/2}}\sum_{\ell}i^{\ell}\left(2\ell+1\right)A_{\ell}\left(r\right)P_{\ell}\left(\cos\theta\right)\nonumber \\
 & \xrightarrow{\mathrm{large\,}r} & \frac{1}{\left(2\pi\right)^{3/2}}\left[e^{i\mathbf{p}\cdot\mathbf{x}}+f\left(\theta\right)\frac{e^{ipr}}{r}\right],
\end{eqnarray}
where the radial partial-wave amplitude $A_{\ell}\left(r\right)$
has the following asymptotic solution for a large $r$:
\begin{equation}
A_{\ell}\left(r\right)=e^{i\delta_{\ell}}\left[\cos\delta_{\ell}\, j_{\ell}\left(pr\right)-\sin\delta_{\ell}\, n_{\ell}\left(pr\right)\right],
\end{equation}
with $j_{\ell}$$\left(n_{\ell}\right)$ as the spherical Bessel (Neumann)
function, and $\delta_{\ell}$ the phase shift for a partial wave
$\ell$ . Thus $f\left(\theta\right)$ can be further expressed as
\begin{equation}
f\left(\theta\right)=\frac{1}{p}\sum_{\ell=0}^{+\infty}\left(2\ell+1\right)e^{i\delta_{\ell}}\sin\delta_{\ell}P_{\ell}\left(\cos\theta\right),\label{eq:scatteringAM}
\end{equation}
with $p=m_{\chi}w/2$. As for the scattering angle $\theta$, to ensure a net capture it
is required to satisfy
\begin{equation}
-c\leq\cos\theta\leq c,
\end{equation}
with
\begin{equation}
c=\frac{v_{c}^{2}-u^{2}-2v_{\mathrm{J}}^{2}}{v_{c}^{2}+u^{2}},
\end{equation}
where $v_{\mathrm{J}}=18.5\,\mathrm{km\cdot s^{-1}}$ is the solar
escape velocity at the radius of Jupiter's orbit. As a conservative
estimate~\citep{Kumar:2012uh}, $v_{\mathrm{J}}$ is introduced not
only to take into account the gravitational effects of the Jupiter
from a more realistic three-body perspective, but to regulate the
divergence encountered in the calculation of $C_{\mathrm{s}}$ for
the long-range self-interaction scenario.

\begin{figure}
\begin{centering}
\includegraphics[scale=0.4]{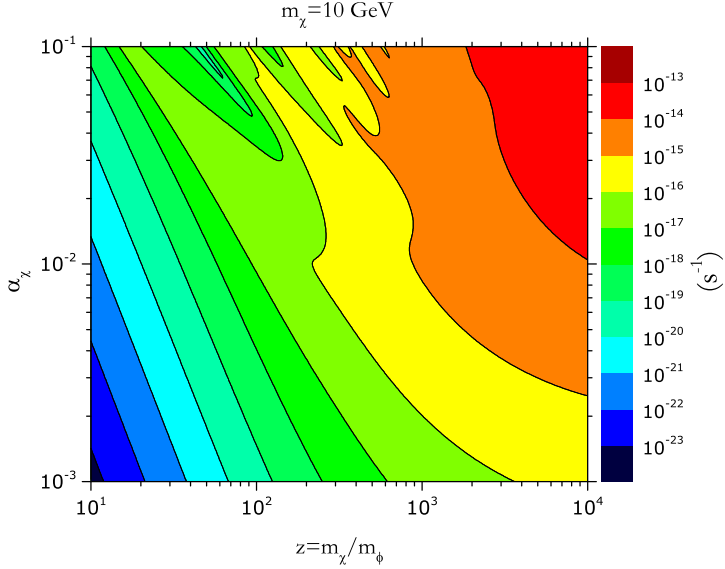}
\par\end{centering}

\protect\caption{\label{fig:Cs0}Shown is the self-capture rate $C_{\mathrm{s}}$ dependent
on parameter $z=m_{\chi}/m_{\phi}$ and $\alpha_{\chi}$ for the Majorana
DM with mass $m_{\chi}=10\,\mathrm{GeV}$.}
\end{figure}

We follow the same procedure as proposed in Ref.~\citep{Tulin:2013teo}
to numerically calculate the phase shift $\delta_{\ell}$ and then
the self-capture rate $C_{\mathrm{s}}$ for specific parameters $m_{\chi}$,
$m_{\phi},$ and coupling strength $\alpha_{\chi}$. The value of
$C_{\mathrm{s}}$ is scanned in Fig.~\ref{fig:Cs0} over the the parameter
space of the mass ratio $z=m_{\chi}/m_{\phi}$ and the self-interaction
strength $\alpha_{\chi}$, for the Majorana DM with mass $m_{\chi}=10\,\mathrm{GeV}$.
While the Born regime occupies
the lower left corner of the figure, the Sommerfeld effect
becomes increasingly significant in the Coulomb limit  ($z\rightarrow\infty$), where a large
partial wave number ($\ell_{\mathrm{max}}\gtrsim1000$) is required
in calculation to ensure the convergence of  $C_{\mathrm{s}}$. It
is worth mentioning that given $z$ and $\alpha_{\chi}$ fixed, $C_{\mathrm{s}}$
will be suppressed for larger DM masses through a $m_{\chi}^{-3}$
dependence, which is obvious from Eq.~(\ref{eq:CsFormula}, \ref{eq:differentialCS},
and~\ref{eq:scatteringAM}).

\subsection{Sommerfeld enhancement of the annihilation rate}

Apart from the self-capture, the light force-carrier brings about significant
consequences on the annihilation of DM as well, which have been widely
studied in determining the thermal relic density and in calculating
signals in indirect DM searches. To gauge the Sommerfeld effect in
annihilation we invoke the Sommerfeld factor as follows~\citep{Cassel:2009wt,Iengo:2009ni}
\begin{eqnarray}
\mathcal{S}_{\ell} & = & \frac{\mathrm{non\mathrm{-}perturbative}\,\mathrm{partial\, wave\, cross\, section}}{\mathrm{tree-level}\,\mathrm{partial\, wave\, cross\, section}}=\left|\frac{\mathcal{M}_{\mathrm{ladder}}^{\ell}}{\mathcal{M}_{0}^{\ell}}\right|^{2},
\end{eqnarray}
where  $\ell$ stands for the  $\ell$-th partial wave. If one skips
somewhat formal theoretical discussions, the implication of Sommerfeld
enhancement on annihilation is encoded in the following heuristic
relation between the non-perturbative $\mathcal{M}_{\mathrm{ladder}}^{\ell}$
and ``bare'' amplitude $\mathcal{M}_{0}^{\ell}$ at tree-level~\citep{Cassel:2009wt,Iengo:2009ni}:
\begin{eqnarray}
\mathcal{M}_{\mathrm{ladder}}^{\ell} & = & \int\mathcal{M}_{0}^{\ell}\left(\left\{ \mathbf{k},\,\mathbf{-k}\right\} \rightarrow\left\{ \mathbf{p}_{f}\right\} \right)\phi\left(\mathbf{k}\right)d^{3}k,
\end{eqnarray}where $\phi\left(\mathbf{k}\right)=\left[\left\langle \mathbf{k}\left|\psi_{\mathbf{p}}^{+}\right.\right\rangle +\left(-1\right)^{S}\left\langle -\mathbf{k}\left|\psi_{\mathbf{p}}^{+}\right.\right\rangle \right]/2$
is the familiar scattering wave function for identical DM-pair in
the reduced system, $\mathbf{p}=m_{\chi}\mathbf{v}_{\mathrm{rel}}/2$
being the initial momentum in the CM frame, with relative velocity
$\mathbf{v}_{\mathrm{rel}}$. It is noted that for identical fermion
pairs $L+S$ must be even, thus we have $\phi\left(\mathbf{k}\right)=\left\langle \mathbf{k}\left|\psi_{\mathbf{p}}^{+}\right.\right\rangle $.
After Fourier transformation and by repeated use of addition theorem
the Sommerfeld factor for partial wave $\ell$ can be written in terms
of the wave function in position space
\begin{eqnarray}
\mathcal{S}_{\ell} & = & \left|\left.\frac{\left(2\ell+1\right)!!}{p^{\ell}\,\ell!}\frac{\partial^{\ell}A_{\ell}\left(r\right)}{\partial r^{\ell}}\right|_{r=0}\right|^{2},
\end{eqnarray}which can be calculated numerically along with $\delta_{\ell}$. Therefore,
after taking into consideration the Sommerfeld enhancement, we express
the annihilation rate for the trapped DM particles as follows
\begin{eqnarray}
C_{\mathrm{a}} & = & \frac{\left\langle \sigma_{\mathrm{ann}}v_{\mathrm{rel}}\right\rangle _{\odot}}{V_{\mathrm{eff}}}\nonumber \\
 & = & \frac{1}{V_{\mathrm{eff}}}\int\left(\sigma_{\mathrm{ann}}\left|\mathbf{v}_{2}-\mathbf{v}_{1}\right|\right)_{\mathrm{tree}}\mathcal{S}\, f_{\chi}\left(\mathbf{v}_{1}\right)f_{\chi}\left(\mathbf{v}_{2}\right)d{}^{3}v_{1}d{}^{3}v_{2}\nonumber \\
 & = & \frac{1}{V_{\mathrm{eff}}}\int\left(\sigma_{\mathrm{ann}}v_{\mathrm{rel}}\right)_{\mathrm{tree}}\mathcal{S}\, f_{\chi}\left(\mathbf{v_{\mathrm{rel}}}/\sqrt{2}\right)d^{3}\left(v_{\mathrm{rel}}/\sqrt{2}\right),\label{eq:annihilationDef}
\end{eqnarray}
where $\mathcal{S}$ is the Sommerfeld factor dependent on $v_{\mathrm{rel}}$,
the footnote ``tree'' indicating that the cross section is calculated
at the tree-level, and the Maxwellian velocity distribution is written as
\begin{equation}
f_{\chi}\left(\mathbf{v}\right)=\frac{e^{-v^{2}/u_{0}^{2}}}{\left(\pi u_{0}^{2}\right)^{3/2}},\label{eq:velocity-distribution}
\end{equation}
with $u_{0}=\sqrt{2T_{\chi}/m_{\chi}}$. Note that in Eq.~(\ref{eq:velocity-distribution})
we neglect the cut-off at the solar escape velocity, which is equivalent to the
approximation $v_{\mathrm{esc}}\rightarrow\infty$.

\begin{figure}
\begin{centering}
\includegraphics[scale=0.29]{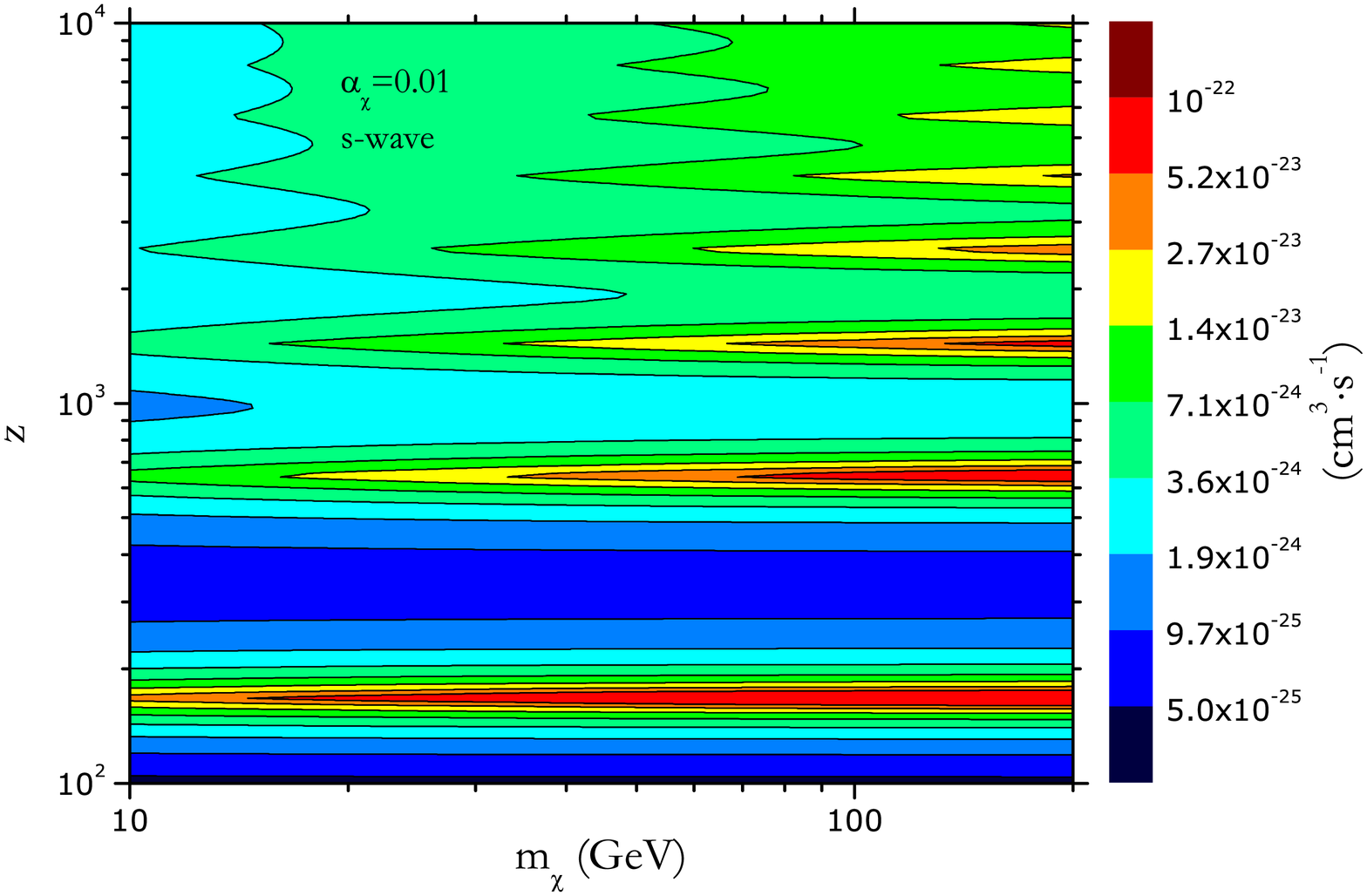}\includegraphics[scale=0.29]{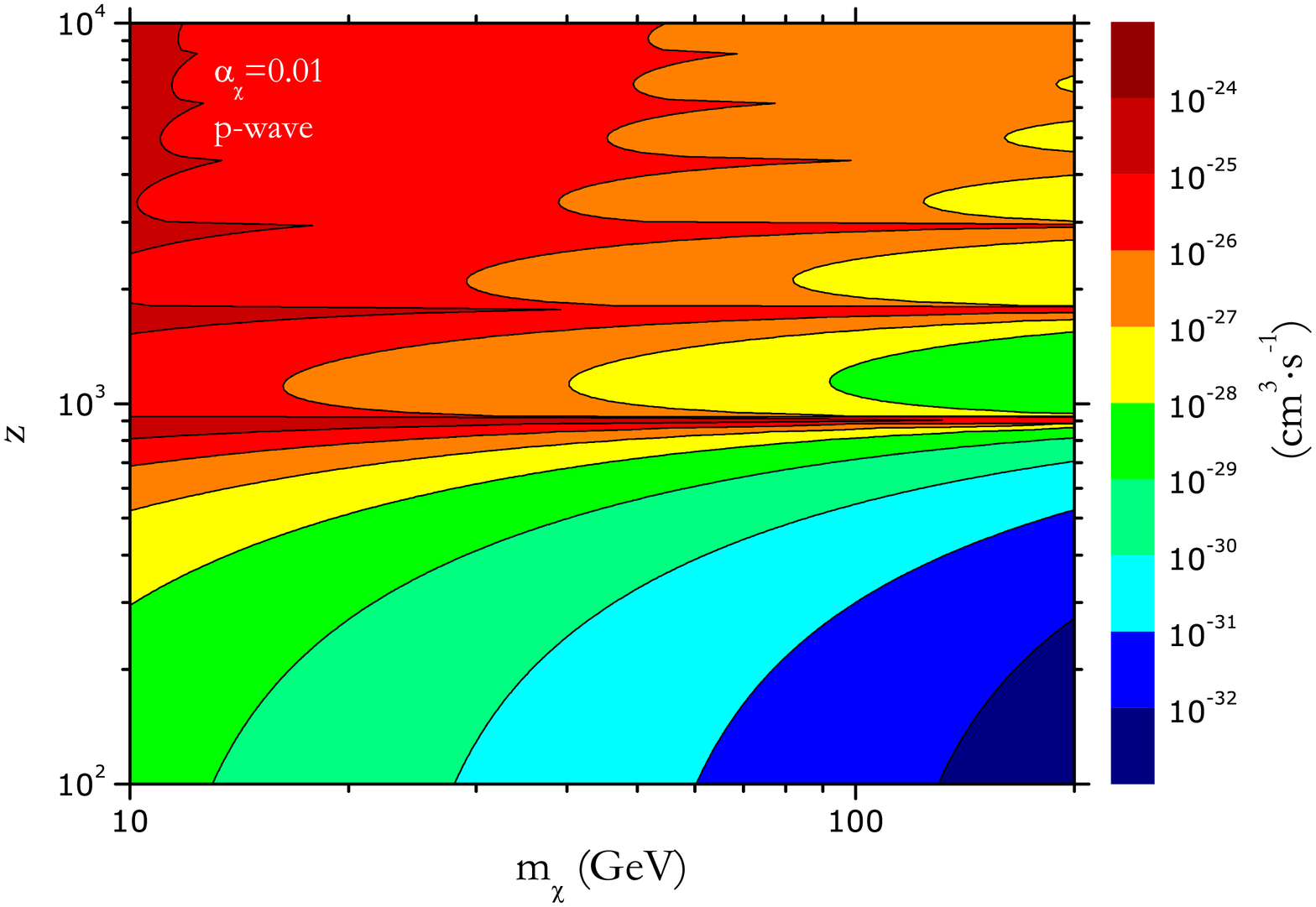}
\par\end{centering}

\protect\caption{\label{fig:SPwave}Shown are the thermally averaged cross section
$\left\langle \sigma_{\mathrm{ann}}v_{\mathrm{rel}}\right\rangle _{\odot}$
with $\alpha_{\chi}=0.01$, for the $s$-wave and $p$-wave annihilation
within the Sun, respectively. See text for details.}
\end{figure}

Since we are only interested in the case where other $p$-wave processes
are suppressed by couplings $y_{f}\sin\theta$ or $y_{N}$ in Eq.~(\ref{eq:model Interacton}), the total thermally
averaged cross section can be written as
\begin{eqnarray}
\left\langle \sigma_{\mathrm{ann}}v_{\mathrm{rel}}\right\rangle  & = & \left\langle \left(\sigma_{\phi\phi}v_{\mathrm{rel}}\right)_{\mathrm{tree}}\mathcal{S}_{1}\right\rangle +\left\langle \sum_{f}\left(\sigma_{\bar{f}f}v_{\mathrm{rel}}\right)_{\mathrm{tree}}\mathcal{S}_{0}\right\rangle \nonumber \\
 & = & \left\langle \left(\sigma_{\phi\phi}v_{\mathrm{rel}}\right)_{\mathrm{tree}}\mathcal{S}_{1}\right\rangle +\left\langle \eta\left(\sigma v_{\mathrm{rel}}\right)_{\mathrm{0}}\mathcal{S}_{0}\right\rangle ,\label{eq:AvergedCrossSection}
\end{eqnarray}
where included are the thermal cross sections for the irreducible
$p$-wave process
\begin{equation}
\left(\sigma_{\phi\phi}v_{\mathrm{rel}}\right)_{\mathrm{tree}}=\frac{3\pi\alpha_{\chi}^{2}v_{\mathrm{rel}}^{2}}{8m_{\chi}^{2}},\label{eq:tree cross section}
\end{equation}
and its $s$-wave counterparts arising from the four-fermion interaction
\begin{eqnarray}
\left(\sigma_{\bar{f}f}v_{\mathrm{rel}}\right)_{\mathrm{tree}} & = & \frac{G_{f}^{2}}{2\pi}c_{f}m_{\chi}^{2}\left(1-\frac{m_{f}^{2}}{m_{\chi}^{2}}\right)^{3/2},
\end{eqnarray}
with the color factor $c_{f}=3\,\left(1\right)$ for quarks (leptons).
$\mathcal{S}_{1}$ $\left(\mathcal{S}_{0}\right)$ is the $p$-wave
($s$-wave) Sommerfeld factor. The $s$-wave component is parametrized
as a small factor $\eta$ times the typical averaged thermal cross
section $\left(\sigma v_{\mathrm{rel}}\right)_{\mathrm{0}}=3\times10^{-26}\,\mathrm{cm^{3}\cdot s^{-1}}$
in the second line of Eq.~(\ref{eq:AvergedCrossSection}). To illustrate
the resonant behaviors of the annihilation, in Fig. \ref{fig:SPwave}
we present the solar thermally averaged cross section $\left\langle \left(\sigma v_{\mathrm{rel}}\right)_{\mathrm{0}}\mathcal{S}_{0}\right\rangle _{\odot}$
and $\left\langle \left(\sigma_{\phi\phi}v_{\mathrm{rel}}\right)_{\mathrm{tree}}\mathcal{S}_{1}\right\rangle _{\odot}$
as the functions of the DM mass $m_{\chi}$ and $z=m_{\chi}/m_{\phi}$,
respectively, with coupling $\alpha_{\chi}$=0.01. The DM mass parameter
$m_{\chi}$ actually reflects the dependence of the DM velocity within
the Sun through the relation $u_{0}=\sqrt{2T_{\chi}/m_{\chi}}$ in
Eq.~(\ref{eq:velocity-distribution}).

\subsection{Constraints from the relic density}

In Sec.\ref{sub:subsectionA} we calculate the self-capture rate $C_{\mathrm{s}}$
with free parameters $\alpha_{\chi}$ and $z=m_{\chi}/m_{\phi}$.
However, the coupling strength $\alpha_{\chi}$ is not an independent
parameter as it can be determined by the observational DM relic density with specific $m_{\chi}$ and $m_{\phi}$. For the given $m_{\chi}$ ,
$m_{\phi}$ and $\alpha_{\chi}$ the evolution of the relic density
of DM is described by the following Boltzmann equation~\citep{Gondolo:1990dk,Kolb:1990vq}
\begin{figure}
\begin{centering}
\includegraphics[scale=0.35]{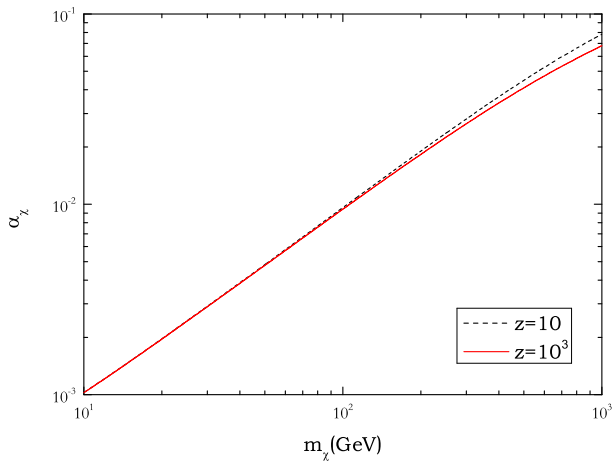}
\par\end{centering}

\protect\caption{\label{fig:alpha}The coupling strength $\alpha_{\chi}$ required
to give the correct relic abundance as a function of DM mass $m_{\chi}$
for $z=10$ (black dashed) and $10^{3}$ (red solid), under the condition
$\eta=10^{-2}$. See text for details.}
\end{figure}
\begin{equation}
\frac{dn_{\chi}}{dt}+3Hn_{\chi}=-\left\langle \sigma_{\mathrm{ann}}v_{\mathrm{rel}}\right\rangle \left(n_{\chi}^{2}-n_{\chi}^{\mathrm{eq}\,2}\right),
\end{equation}
where $n_{\chi}$ and $n_{\chi}^{\mathrm{eq}}$ are the number density
of the DM and its counterpart in equilibrium with the background particles,
respectively, $H$ the Hubble constant and $\left\langle \sigma_{\mathrm{ann}}v_{\mathrm{rel}}\right\rangle $
is the thermally averaged cross section introduced in Eq.~(\ref{eq:AvergedCrossSection}).
In practice by changing variables such that $Y=n_{\chi}/s\,\left(Y^{\mathrm{eq}}=n_{\chi}^{\mathrm{eq}}/s\right)$
with $s$ the entropy density and $x=m_{\chi}/T$, the above Boltzmann
equation can be rewritten as follows \citep{Gondolo:1990dk,Kolb:1990vq}
\begin{equation}
\frac{dY}{dx}=-\sqrt{\frac{\pi}{45}}m_{\mathrm{P}}m_{\chi}\frac{\left(g_{*s}/\sqrt{g_{*}}\right)}{x^{2}}\left\langle \sigma_{\mathrm{ann}}v_{\mathrm{rel}}\right\rangle \left(Y^{2}-Y^{\mathrm{eq}\,2}\right),\label{eq:BoltzmannEquation}
\end{equation}
where $g_{*s}$ and $g_{*}$ are the effective degrees of freedom
for entropy and energy density, respectively, and the Planck mass
$m_{\mathrm{P}}=1.22\times10^{19}\,\mathrm{GeV}$. We follow Ref.
\citep{2011JCAP...01..016H,Chen:2013bi} to numerically solve Eq.~(\ref{eq:BoltzmannEquation}) so as to give the coupling $\alpha_{\chi}$
determined by the relic density. In Fig.~\ref{fig:alpha} shown is
the value of $\alpha_{\chi}$ as a function of $m_{\chi}$ and $z=m_{\chi}/m_{\phi}$,
with a specific value $\eta=10^{-2}$ introduced in Eq.~(\ref{eq:AvergedCrossSection}). One can see from Fig.~\ref{fig:alpha} that owing to the high velocity at the chemical decoupling stage, $\alpha_{\chi}$
is insensitive to the  Sommerfeld effect.
Here we note that the effect of the kinetic decoupling is not included
in our calculation, because it is insignificant for the case of $p$-wave-dominated
annihilation when the temperature of kinetic decoupling $T_{\mathrm{kd}}$
is much less than that of the chemical decoupling $T_{f}$, even if
the Sommerfeld enhancement is taken into consideration \citep{2010PhLB..687..275D,Chen:2013bi,Tulin:2013teo}.
This requirement is in coincidence with the onset of the Sommerfeld
effect: a large $z=m_{\chi}/m_{\phi}\gg1$ guarantees that $T_{f}\simeq m_{\chi}/25\gg m_{\phi}\sim T_{\mathrm{kd}}^{\phi}$,
where the kinetic decoupling temperature for the $\chi$-$\phi$ interaction
$T_{\mathrm{kd}}^{\phi}$ provides a maximal value for $T_{\mathrm{kd}}$~\citep{Feng:2010zp}.  Given $\eta=0.01$,
in Fig.~\ref{fig:SolarP-S-Wave} we use the relic density to calculate
the thermally averaged cross section for the $s$-wave $\left\langle \left(\sigma v_{\mathrm{rel}}\right)_{\mathrm{0}}\mathcal{S}_{0}\right\rangle _{\odot}$
and $p$-wave annihilation $\left\langle \left(\sigma_{\phi\phi}v_{\mathrm{rel}}\right)_{\mathrm{tree}}\mathcal{S}_{1}\right\rangle _{\odot}$
within the Sun, respectively. We note that smaller $\eta$ will not
bring remarkable changes to Fig. \ref{fig:SolarP-S-Wave} because
the relic density turns out to remain constant as the $s$-wave component
gets smaller ($\eta<0.01$) \citep{Chen:2013bi}.
\begin{figure}
\begin{centering}
\includegraphics[scale=0.29]{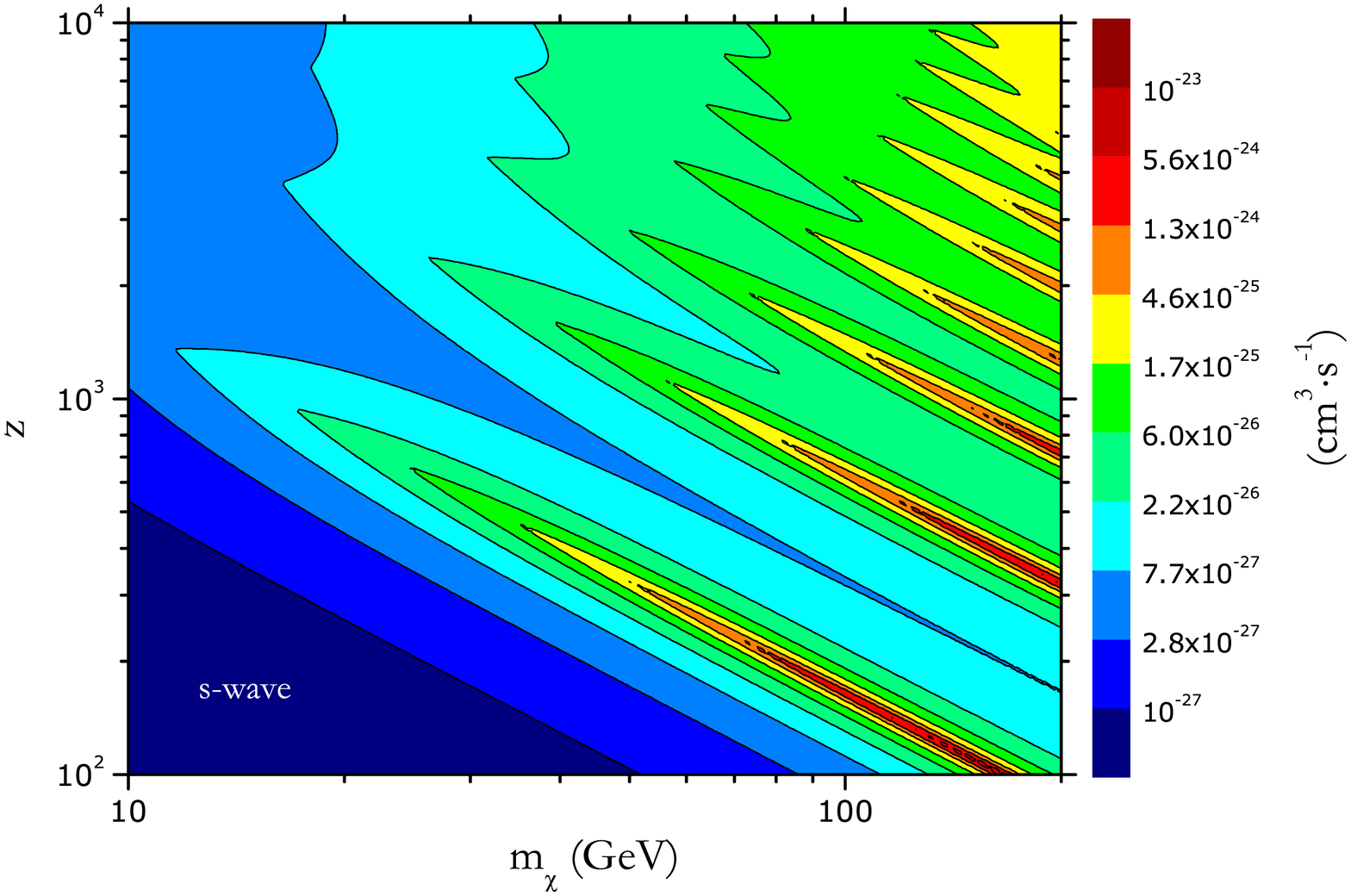}\includegraphics[scale=0.29]{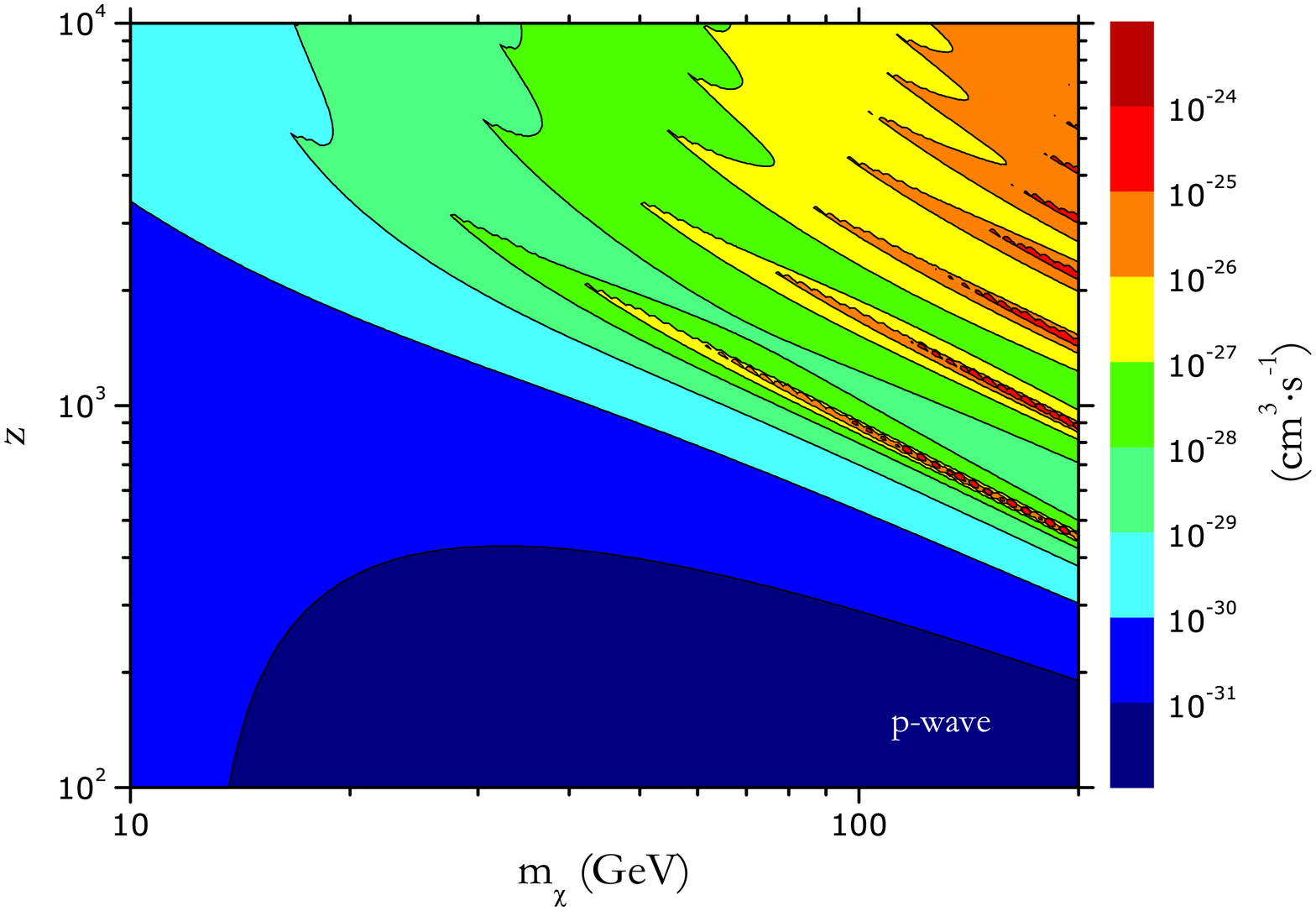}
\par\end{centering}

\protect\caption{\label{fig:SolarP-S-Wave} The solar thermal cross section
$\left\langle \sigma_{\mathrm{ann}}v_{\mathrm{rel}}\right\rangle _{\odot}$
determined by the relic density for the $s$-wave and $p$-wave
annihilation, respectively. See text for details. }
\end{figure}

\section{Consequences on the neutrino telescope\label{sec:4}}

The signal of  the solar DM is associated with the observation of
primary and secondary high energy neutrinos, the annihilation products of the
DM accumulated in the center of the Sun. Among other goals, many terrestrial
neutrino detection projects such as IceCube \citep{IceCubecollaboration2012},
Super-Kamiokande~\citep{2011ApJ...742...78T}, Baikal Neutrino Project~\citep{Avrorin:2014swy} and ANTARES~\citep{Adrian-Martinez:2013ayv}
are dedicated to such observation, and next generation telescopes
including PINGU~\citep{Aartsen:2014oha} and Hyper-Kamiokande~\citep{Abe:2011ts}
are also in proposal. In general the neutrino differential flux at the
detector location can be schematically described as
\begin{equation}
\frac{d\Phi_{\nu}}{dE_{\nu}}=\frac{\Gamma_{A}}{4\pi d_{\odot}^{2}}\frac{dN_{\nu}}{dE_{\nu}},
\end{equation}
where $\Gamma_{A}$ is the annihilation rate introduced in Eq.~(\ref{eq:annihilation rate}),
$d_{\odot}$ the Sun-Earth distance, and $dN_{\nu}/dE_{\nu}$ denotes
the energy spectrum of neutrino $\nu_{e,\mu,\tau}$ and $\bar{\nu}_{e,\mu,\tau}$
per DM annihilation event, after folding the effects of annihilation
branch ratio, hadronization of quarks and neutrino oscillation, which
are usually simulated with $\mathtt{DarkSUSY}$ \citep{Gondolo:2004sc}
and $\mathtt{WimpSim}$ \citep{Blennow:2007tw}. For the purpose of
illustration the neutrino flux is connected to the readout at the
detector terminal with the following relation:

\begin{equation}
N_{\mu}=\int_{0}^{t_{\mathrm{exp}}}dt\int_{\Delta\varOmega}d\Omega\int_{E_{\mathrm{th}}}\frac{d\Phi_{\nu}}{dE_{\nu}}A_{\nu}\left(E_{\nu}\right)dE_{\nu},
\end{equation}
where $N_{\mu}$ is the number of the muon events that arise from
the charge and neutral current interaction between the incident neutrino
and medium surrounding the detector, the effects of which is encoded
in the effective area $A_{\nu}$, $E_{\mathrm{th}}$ the threshold
energy of the detector, $\Delta\varOmega=2\pi\left(1-\cos\Psi\right)$
being the solid angle surrounding the Sun with angular resolution
$\Psi$, and $t_{\mathrm{exp}}$ is the exposure time. If no excess
of muon over the background is observed, an upper limit on the annihilation
rate $\Gamma_{A}$ can be obtained.

Now we use the recent release of the Super-Kamiokande (SK) analysis~\citep{Choi:2015ara} to constrain the strength of the DM-nucleon
interaction in the presence of the Sommerfeld effect, in terms of
the upper limits on the nuclear capture rate $C_{\mathrm{n}}$ that
depends on the specific DM-nucleon interaction model. One can translate
these constraints back on the couplings of the DM-nucleon interaction
with the nuclear form factors given by Ref.~\citep{Catena:2015uha,Catena:2015iea},
once the interaction is specified. For the purpose of simple illustration
we assume a leptophilic four-fermion interaction in Eq.~(\ref{eq:model Interacton}),
so that the incoming DM particle with velocity $w$ interacts with
the target nucleus through the Higgs portal. The relevant differential
cross section is expressed as
\begin{eqnarray}
\frac{d\sigma_{\chi N}}{dq^{2}} & = & \alpha_{\chi}f_{N}^{2}\sin^{2}\theta\left(\frac{m_{p}}{v_{\mathrm{EW}}}\right)^{2}\frac{A^{2}F_{N}^{2}(q^{2})}{\left(q^{2}+m_{\phi}^{2}\right)^{2}}\frac{1}{w^{2}},\label{eq:nuclearCrossSection}
\end{eqnarray}
where $q$ is the transferred momentum of the incident DM particle,
and $A$ is the atomic number of the target nucleus. The nucleon mass
is approximated as the protonic one $m_{p}$, and the coefficient
$f_{N}=0.351$ arises from the scalar bilinear for the underlying
quarks. $F_{N}^{2}(q^{2})=\left[3\,\mathrm{j_{1}}\left(q\, R_{1}\right)/\left(q\, R_{1}\right)\right]^{2}e^{-q^{2}s^{2}}$
is the nuclear form factor, where $\mathrm{j_{1}}(x)=\sin\left(x\right)/x^{2}-\cos\left(x\right)/x$
is the Bessel spherical function of the first kind, and the parameter
$R_{1}=\sqrt{R_{0}^{2}-5s^{2}}$ with $R_{0}\backsimeq1.23A^{1/3}\,\mathrm{fm}$,
and $s\backsimeq1\,\mathrm{fm}$~\citep{Lewin:1995rx}.

\begin{figure}
\begin{centering}
\includegraphics[scale=0.305]{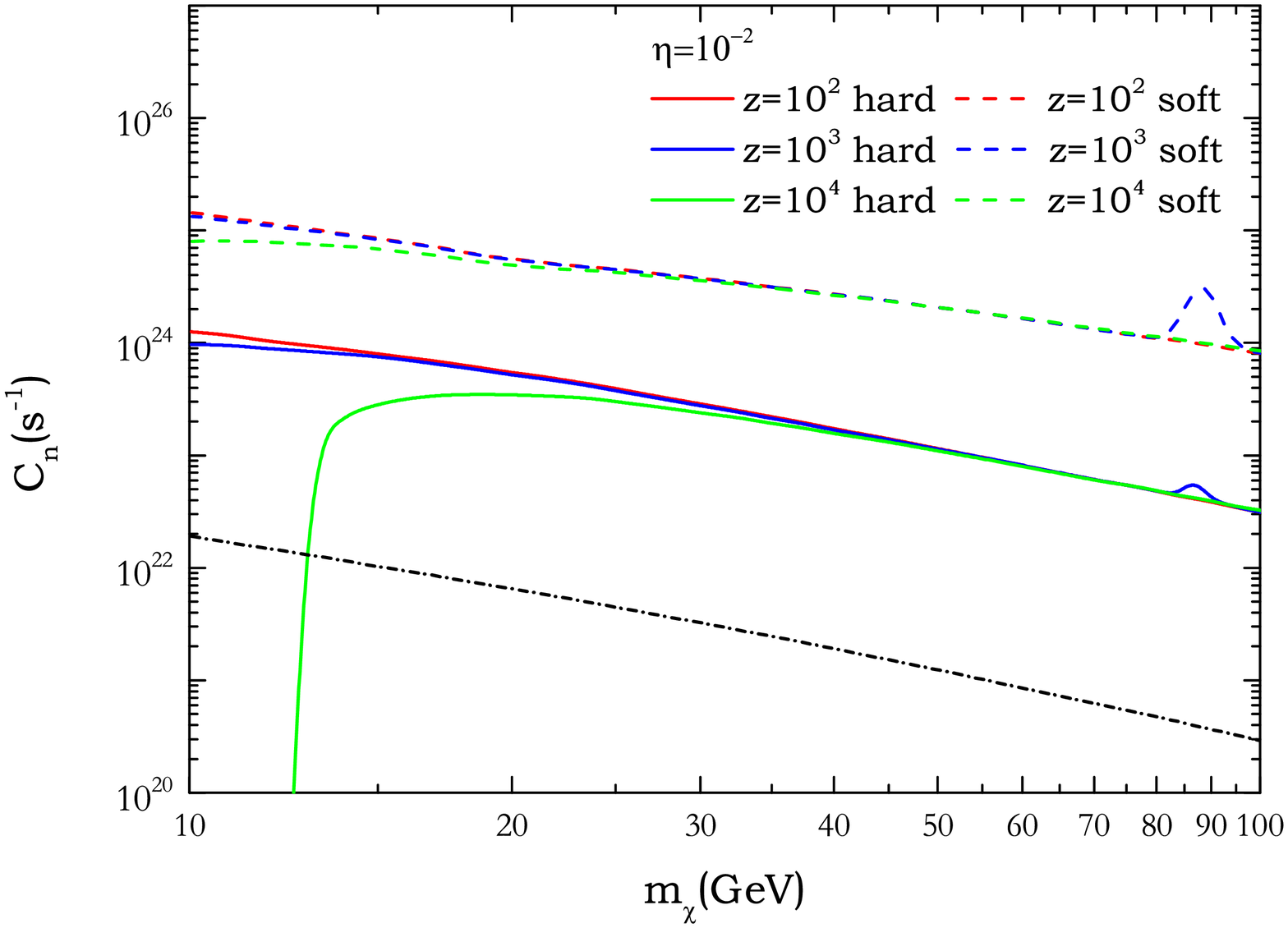}\includegraphics[scale=0.305]{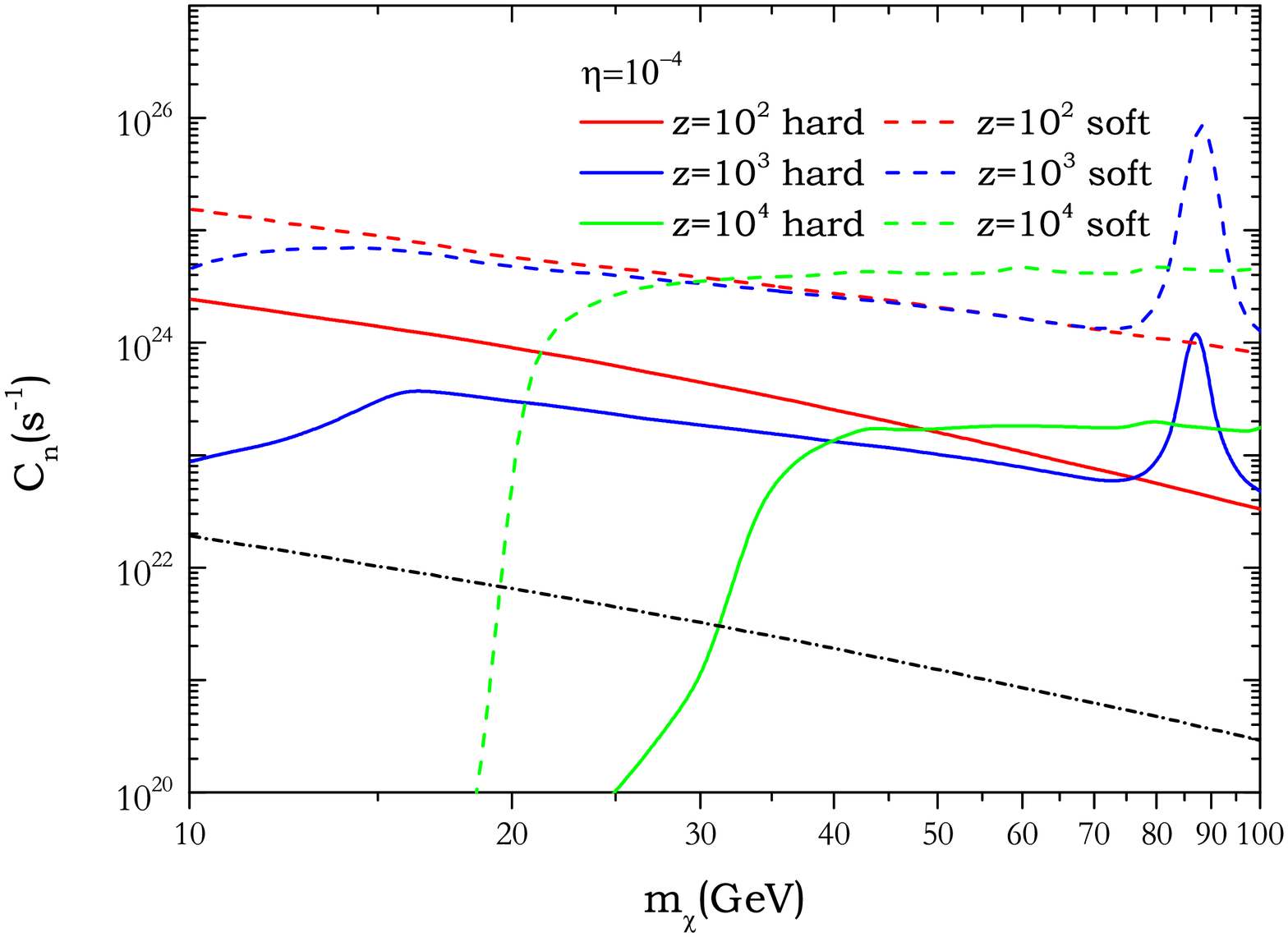}
\par\end{centering}

\begin{centering}
\includegraphics[scale=0.305]{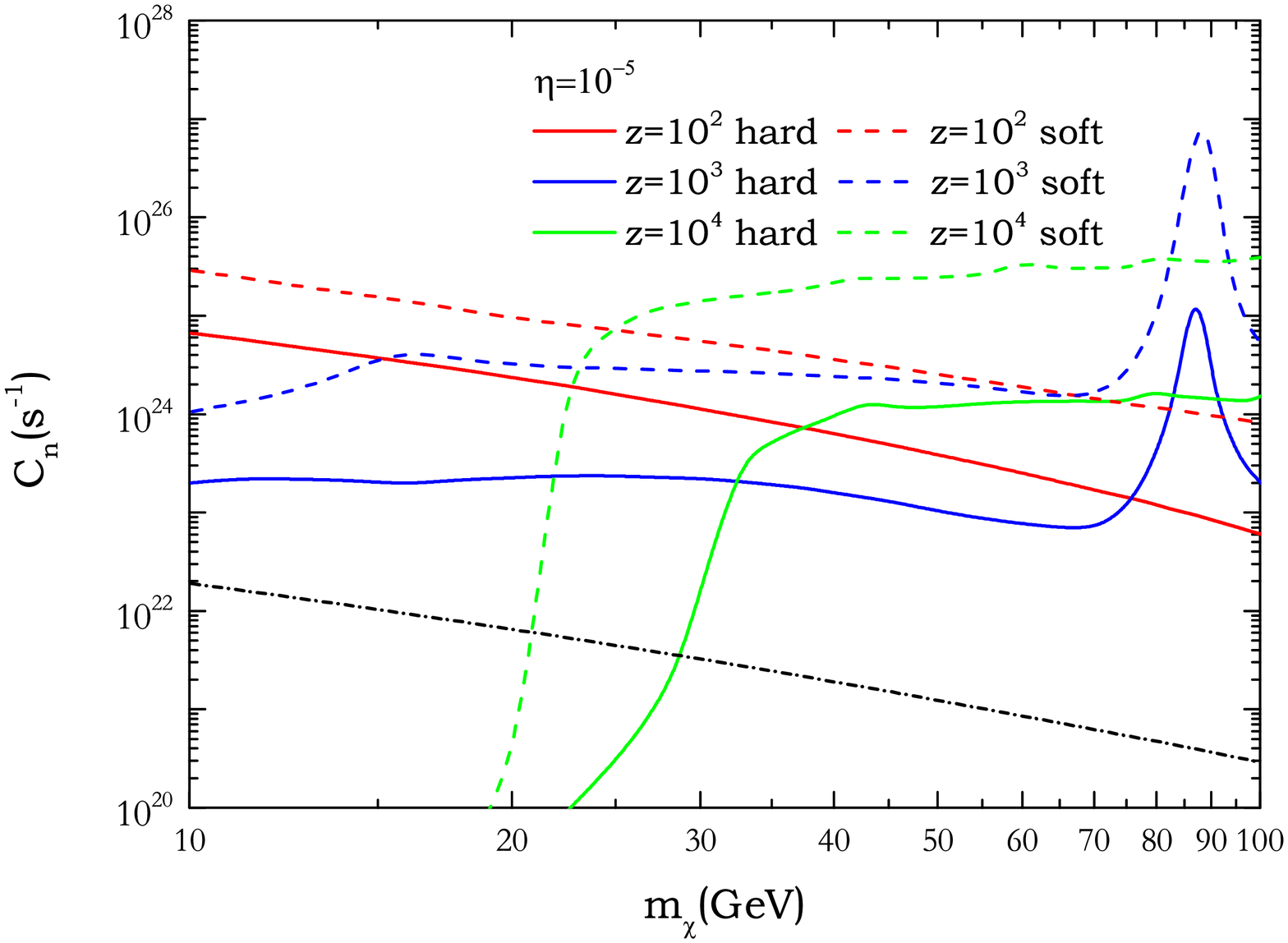}\includegraphics[scale=0.305]{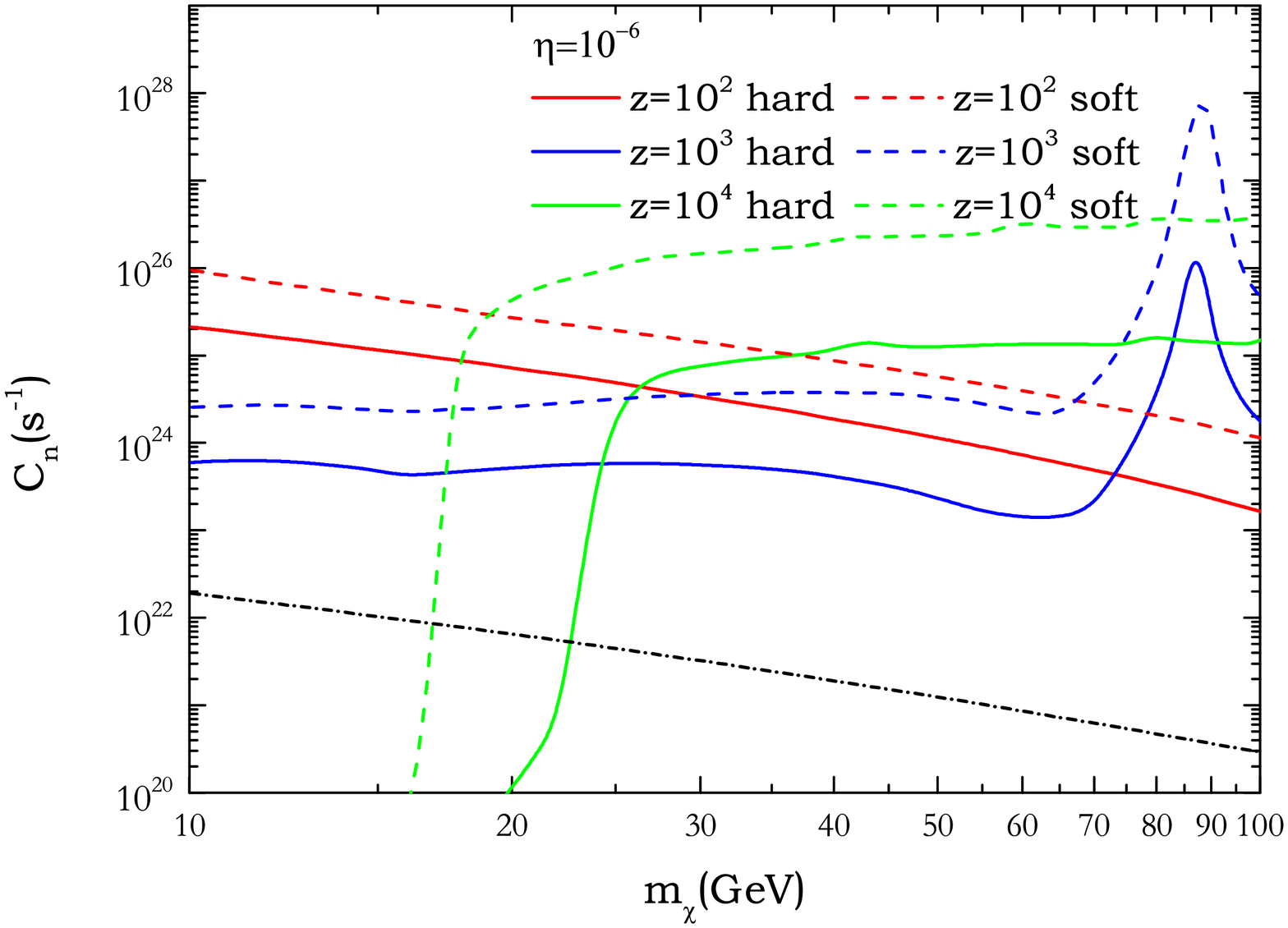}
\par\end{centering}

\protect\caption{\label{fig:Cn} SK 90\% C. L. upper limits on nuclear capture rate
$C_{\mathrm{n}}$ over a range of DM masses for $\eta=10^{-2},\,10^{-4},\,10^{-5}\,\mathrm{and}\,10^{-6}$,
respectively. Solid (dashed) lines indicate hard (soft) annihilation
channels. Shown in the dash-dot line is the nuclear capture rate calculated
with parameters $\sin\theta=10^{-8}$ and $z=10^{4}$. See text for
details.}
\end{figure}

Current DM direct detections provide a stringent upper bound on the
mixing angle such that $\sin\theta\lesssim10^{-8}$~\citep{Kouvaris:2014uoa}.
Mixing angle at such scale is 2 to 3 orders of magnitude smaller for
the light force-carrier $\phi$ to decay before $t\sim1\,\mathrm{s}$,
jeopardizing the success of the BBN, so $\phi$ is assumed to
decay quickly to the sterile neutrino $N$ instead~\citep{Kouvaris:2014uoa}, as
explained below Eq.~(\ref{eq:model Interacton}). As a consequence
we regard the $s$-wave annihilation as the only source relevant for
the DM indirect detection.  In Fig.~\ref{fig:Cn} we present the
upper limits at 90\% confidence level (C.~L) on $C_{\mathrm{n}}$
imposed by SK with various $z=m_{\chi}/m_{\phi}=10^{2},\,10^{3},\,\mathrm{and}\,10^{4}$,
and $\eta=10^{-2},\,10^{-4},\,10^{-5}\,\mathrm{and}\,10^{-6}$, respectively.
As is discussed similarly in Ref.~\citep{Choi:2015ara}, two extreme
benchmark scenarios in which DM exclusively annihilate via hard $\left(\tau^{+}\tau^{-}\right)$
and soft $\left(b\bar{b}\right)$ $s$-wave channel are taken into
consideration over a DM mass range from $10\,$ to $100\,\mathrm{GeV}$. The number of the trapped solar DM particles $N_{\chi}$, or equivalently,
the nuclear capture rate $C_{\mathrm{n}}$ is constrained by the upper
limit on the annihilation rate $\varGamma_{\mathrm{up}}$ with the
following relation:
\begin{equation}
\varGamma_{\mathrm{up}}\geq\frac{1}{2}C_{a}N_{\chi}^{2}\,\mathrm{BR}_{s},
\end{equation}
where the branch ratio of the $s$-wave annihilation which is responsible
for the neutrino signal can be expressed as
\begin{equation}
\mathrm{BR}_{s}=\left\langle \eta\left(\sigma v_{\mathrm{rel}}\right)_{\mathrm{0}}\mathcal{S}_{0}\right\rangle _{\odot}/\left[\left\langle \left(\sigma_{\phi\phi}v_{\mathrm{rel}}\right)_{\mathrm{tree}}\mathcal{S}_{1}\right\rangle _{\odot}+\left\langle \eta\left(\sigma v_{\mathrm{rel}}\right)_{\mathrm{0}}\mathcal{S}_{0}\right\rangle _{\odot}\right].
\end{equation}

The constraints on $C_{\mathrm{n}}$ are affected mainly by three factors:
$\varGamma_{\mathrm{up}}$, $z$ and $\eta$. As a rough estimate,
Eq.~(\ref{eq:annihilation rate}) can be used to explain the features
of the constraints in Fig.~\ref{fig:Cn}: when the Sommerfeld effect
is insignificant ($z=10^{2},\,10^{3}$) so that $\varGamma_{\mathrm{up}}\gg C_{\mathrm{s}}^{2}/\left(2C_{a}\right)$,
upper limits are sensitive to $\eta$ through the dependence on
$\mathrm{BR}_{s}$; the sharp increases of the sensitivity on $C_{\mathrm{n}}$
in Fig.~\ref{fig:Cn} imply that $\varGamma_{\mathrm{up}}\sim\mathcal{O}\left(C_{\mathrm{s}}^{2}\,\mathrm{BR}_{s}/\left(2C_{a}\right)\right)$
when the Sommerfeld effect turns remarkable ($z=10^{4}$), which is of
our interest in this study. For reference purpose in Fig.~\ref{fig:Cn}
we also present the nuclear capture rate (in the black dash-dot line)
derived from the DM-nucleus cross section in Eq.~(\ref{eq:nuclearCrossSection}),
for which the values of the $\phi$-Higgs mixing angle $\sin\theta=10^{-8}$
and $z=10^{4}$ are adopted. Parameters at such scales are comparable
to the sensitivity of direct detection discussed in Ref.~\citep{Kouvaris:2014uoa}. Although the direct detection approaches are proved to be efficient
in exploring the spin-independent nucleon-DM interaction, the solar
neutrino detection may impose even more stringent constraints on that as the relevant parameter space is significantly
squeezed by the Sommerfeld effect. Such phenomenon occurs at $z=10^{4}$
for the present SK sensitivity, as is shown in Fig.~\ref{fig:Cn}.
\begin{figure}
\begin{centering}
\includegraphics[scale=0.3]{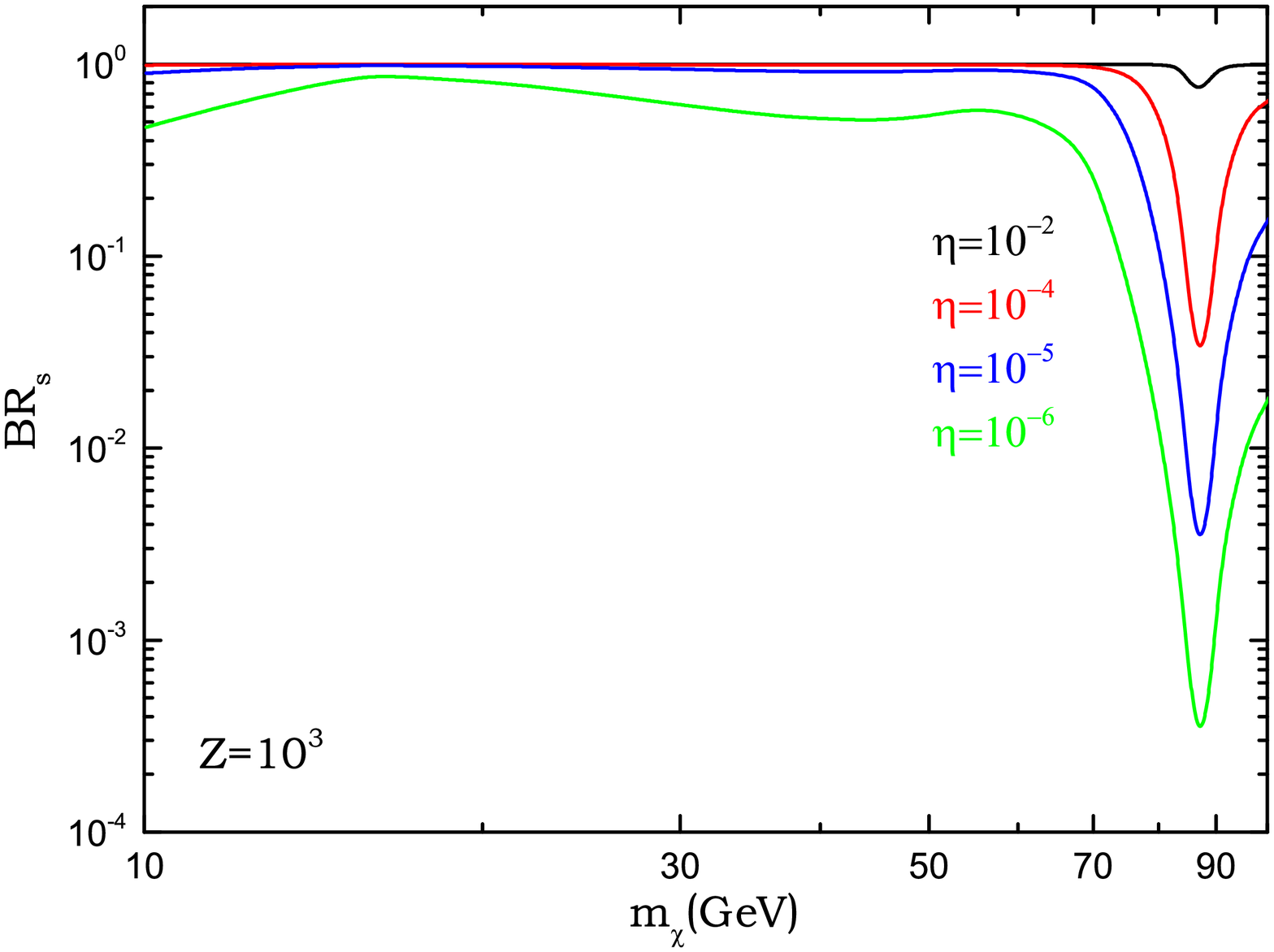}\includegraphics[scale=0.3]{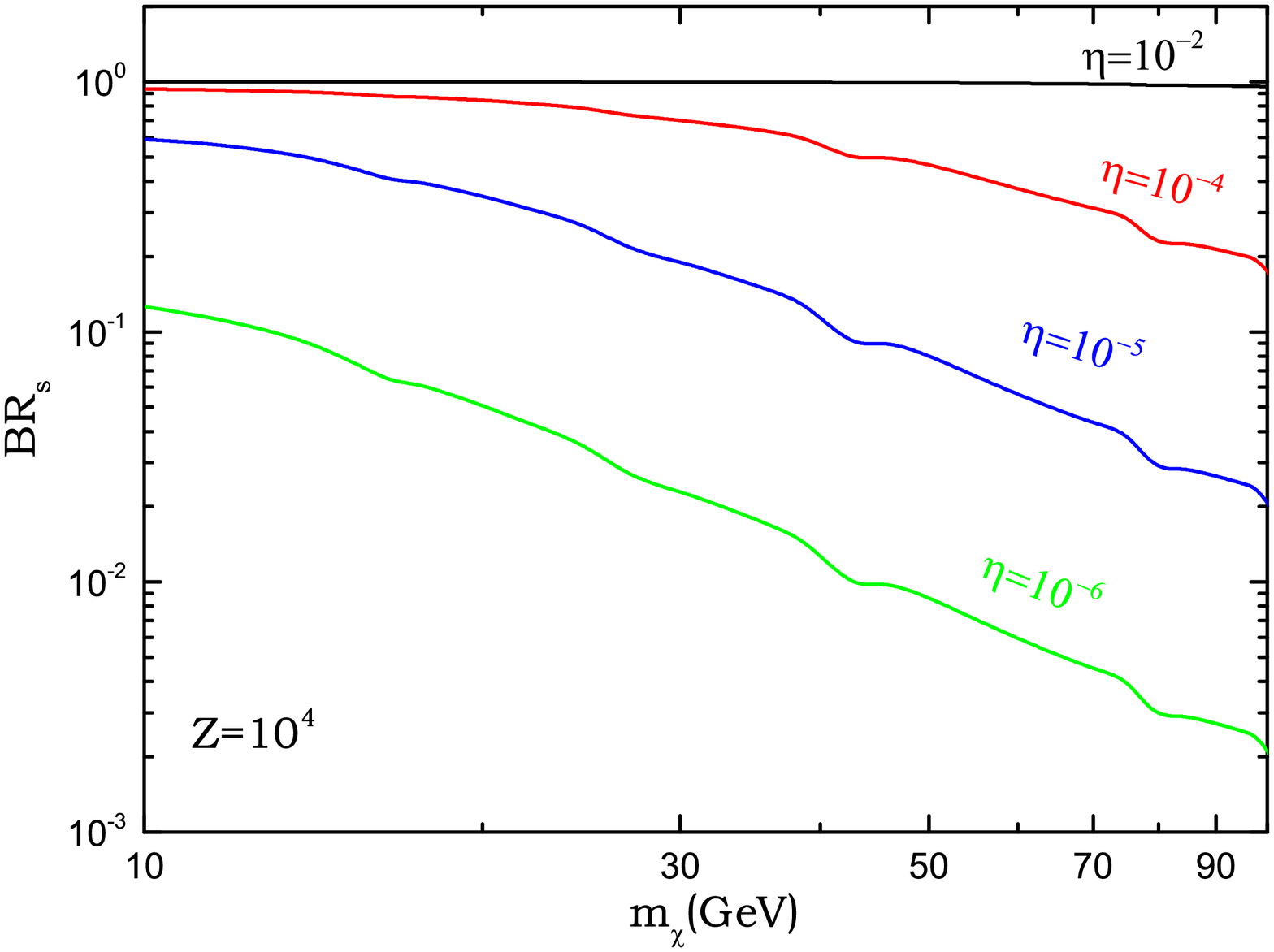}
\par\end{centering}

\protect\caption{\label{fig:branchratio} For the given $z=10^{3}$ and $z=10^{4}$,
branch ratio of the $s$-wave channel as the function of $m_{\chi}$
in the Sun. }
\end{figure}

It is also noted that when the $s$-wave component is sufficiently
small that its cross section drops below the $p$-wave counterpart,
constraints will turn loose because that's when $\mathrm{BR}_{s}$
begins to be suppressed by a small $\eta$. To get some sense, in
Fig.~\ref{fig:branchratio} we plot $\mathrm{BR}_{s}$ as a function
of DM mass $m_{\chi}$ for $z=10^{3}$ and $z=10^{4}$. It is evident
that the dips in $\mathrm{BR}_{s}$ in the left panel are attributed to the
$p$-wave resonance encountered along the cross section at $z=10^{3}$
in Fig.~\ref{fig:SolarP-S-Wave}, and hence are responsible for the
corresponding bumps in sensitivity on $C_{\mathrm{n}}$ in Fig.~\ref{fig:Cn}
for $z=10^{3}$.

\section{Conclusions\label{sec:5}}

In this work we investigate an example of the solar DM that self-interact
via light force carriers, specifically, Majorana DM particles scattering with long-range self-interaction and annihilating dominantly through $p$-wave
channel. To optimize and cover the relevant phenomenology of the solar
SIDM as comprehensively as possible, we study a scenario where a small but
non-vanishing quota of DM annihilation proceeded through the $s$-wave
channel in the thermal decoupling epoch, so that the self-capture,
both $s$- and $p$-wave annihilation are all involved in our consideration.

Instead of evaluating relevant DM self-capture and annihilation processes
at tree level, we calculate them in a non-perturbative way by solving
the  Schrödinger  equation with Yukawa potential. We find that the Sommerfeld effect can be significant for   DM self-capture if
coupling $\alpha_{\chi}$  and mass ratio $z$  are regarded as  free parameters. Besides,
the impact of the light mediator on the thermal freeze-out
that is assumed to produce the the observed DM relic density is also
taken into consideration, and is used to determine the self-interaction
coupling $\alpha_{\chi}$ for the given masses of DM particle and mediator.

Moreover we also explore the consequence of the Sommerfeld effect on
terrestrial neutrino telescopes which is utilized to probe signals
from the solar DM. Since the light mediators produced from the $p$-wave
DM annihilation do not produce any observable signal at the terrestrial
neutrino telescope, the neutrino signals originate exclusively
from the $s$-wave component, which may become close to or even dominant
over its $p$-wave counterpart within the Sun, due to the Sommerfeld
enhancement. Therefore we combine the sensitivity of SK and the observed
DM relic density to constrain the nuclear capture rate, or equivalently,
the strength of DM-nucleon interaction. Due to SK's high sensitivity,
the constraint on the DM-nucleon interaction can be remarkably tightened
in the low DM mass region, where the Sommerfeld effect on the DM self-capture
rate has not yet been outstripped by the $m_{\chi}^{-3}$ suppression.

It is also noted that while our discussion is based on a generic class
of models in which the fermion DM particles self-interact via a light
$scalar$  mediator, the consequence of this force-carrier in
astrophysics and cosmology is another interesting topic that is beyond
the scope of this work (For a most recent study, see Ref.~\citep{Kainulainen:2015sva}). Besides, as a tentative study on the phenomenology of the solar SIDM, our discussion can be further extended to
other SIDM models, not confined to the case of scalar mediator or
Majorana DM.

Simple as it is, this work has captured the main features
of the solar SIDM. To put it in a nutshell, the interplay between the long-range
dark force and the solar neutrino detection is encoded in the following
expression up to a possible correction due to the self-capture saturation,
\begin{eqnarray}
\varGamma_{\mathrm{up}} & \geq & \frac{1}{2}\left[C_{\mathrm{n}}+\frac{1}{2}\frac{C_{\mathrm{s}}^{2}\, V_{\mathrm{eff}}}{\left(\left\langle \sigma v_{\mathrm{rel}}\right\rangle _{s\mathrm{wave}}+\left\langle \sigma v_{\mathrm{rel}}\right\rangle _{p\mathrm{wave}}\right)}+\cdots\right]\mathrm{BR_{signal}},
\end{eqnarray}
where $\left\langle \sigma v_{\mathrm{rel}}\right\rangle _{s\mathrm{wave}}$
($\left\langle \sigma v_{\mathrm{rel}}\right\rangle _{p\mathrm{wave}}$)
is the $s$-wave ($p$-wave) thermal cross section, and $\mathrm{BR_{signal}}$
is the branch ratio for the signal annihilation channel. Qualitatively
speaking, a model with a combination of enhanced $C_{\mathrm{s}}$
and velocity-suppressed $\left\langle \sigma v_{\mathrm{rel}}\right\rangle _{p\mathrm{wave}}$
may maximize the novel effects of the SIDM, as long as such enhancement
has not been offset by an extremely small $\mathrm{BR_{signal}}$.
For instance, in the model where $\phi$ decays to the active neutrinos
before the BBN~\citep{Kouvaris:2014uoa}, $\mathrm{BR_{signal}}$
will be significantly boosted if such decay proceeds before $\phi$
reaches the Earth. In this case, even for a smaller $z$ the neutrino
detector SK will lead in the sensitivity on $C_{\mathrm{n}}$.
\appendix

\begin{acknowledgments}
We would like to thank Yang Ziqing for useful discussion. This work is supported in part by the National Basic Research Program of China (973 Program) under Grants No. 2010CB833000; the National Nature Science Foundation of China (NSFC) under Grants No. 10905084, No. 11335012 and No. 11475237; The numerical calculations were done using the HPC Cluster of SKLTP/ITP-CAS.
\end{acknowledgments}

\section{\label{sec:appendixDMDistribution}Distribution of the self-interacting
dark matter in the sun}
The distribution of the DM particles trapped in the Sun has been well
studied in the literature~\citep{Spergel:1984re,Nauenberg:1986em,Gould:1987ju}.
While in Ref.~\citep{Spergel:1984re} authors adopts a Maxwellian velocity distribution
$f_{\chi}\left(\mathbf{v}\right)\propto\exp\left(-E_{\mathrm{k}}/T_{\chi}\right)\equiv\exp\left[-m_{\chi}v^{2}/\left(2\, T_{\chi}\right)\right]$,\footnote[1]{\renewcommand{\baselinestretch}{1}\selectfont  By demand of satisfying the  Boltzmann equation, the Maxwellian  velocity distribution in turn leads to  the Boltzmann distribution $f_{\chi}\left(\mathbf{r},\,\mathbf{v}\right)\propto\exp\left[-\left(V\left(r\right)+E_{\mathrm{k}}\right)/T_{\chi}\right]$,
and hence the density distribution $n_{\chi}\left(\mathbf{r}\right)\propto\exp\left[-V\left(r\right)/T_{\chi}\right]$, with $V\left(r\right)$  the gravitational potential of
the DM particle at  radius $r$.}
to describe the non-thermal equilibrium between the DM particles and solar nuclei, the studies in Ref.
\citep{Nauenberg:1986em,Gould:1987ju} indicate a deviation from the
Maxwellian form, in a manner that the actual velocity distribution
is suppressed at the tail and tends to be anisotropic at large radius.
Such deviation can be attributed to the fact that the energetic collisions
that send the DM particles into the high orbits occur predominantly near
the hotter core of the Sun. Thus one expects a lower angular
momentum distribution, or equivalently a larger radial-velocity component
$\left(v_{r}/v\right)$ for the high-energy orbits. This is illustrated
in Fig.~2  of  Ref.~\citep{Gould:1987ju}, where the normalized average
kinetic energy of the DM particles $\left(2/3\right)\left\langle E_{\mathrm{k}}\right\rangle /T_{\chi}$
decreases toward the outer layers of the Sun, and the average square
of the radial velocity component $\left\langle \left(v_{r}/v\right)^{2}\right\rangle $
deviates from $1/3$ at radii larger than around $0.2\, R_{\odot}$,
indicating an anisotropic nature of the velocity distribution beyond that radius. Nevertheless, it should be stressed that while such correction to
the thermal distribution is necessary  for the evaluation of
the energy transport and evaporation in the Sun that rely sensitively
on the high end of the velocity distribution~\citep{Nauenberg:1986em,Gould:1987ju},
the phenomenology of interest in this study such as self-capture
and annihilation can still be described with the bulk of the Boltzmann form to a good approximation.

\begin{figure}
\begin{centering}
\includegraphics[scale=0.275]{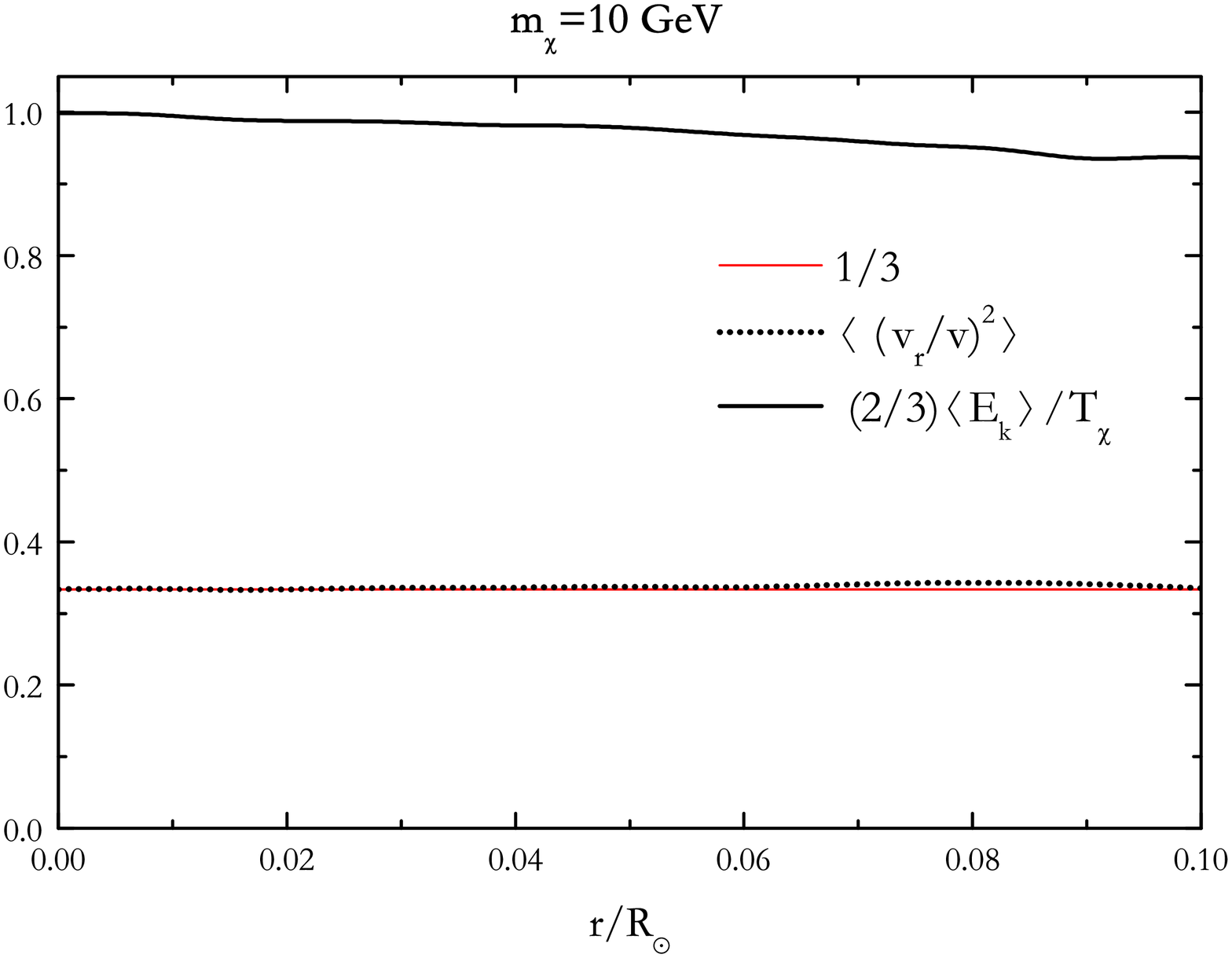}\includegraphics[scale=0.275]{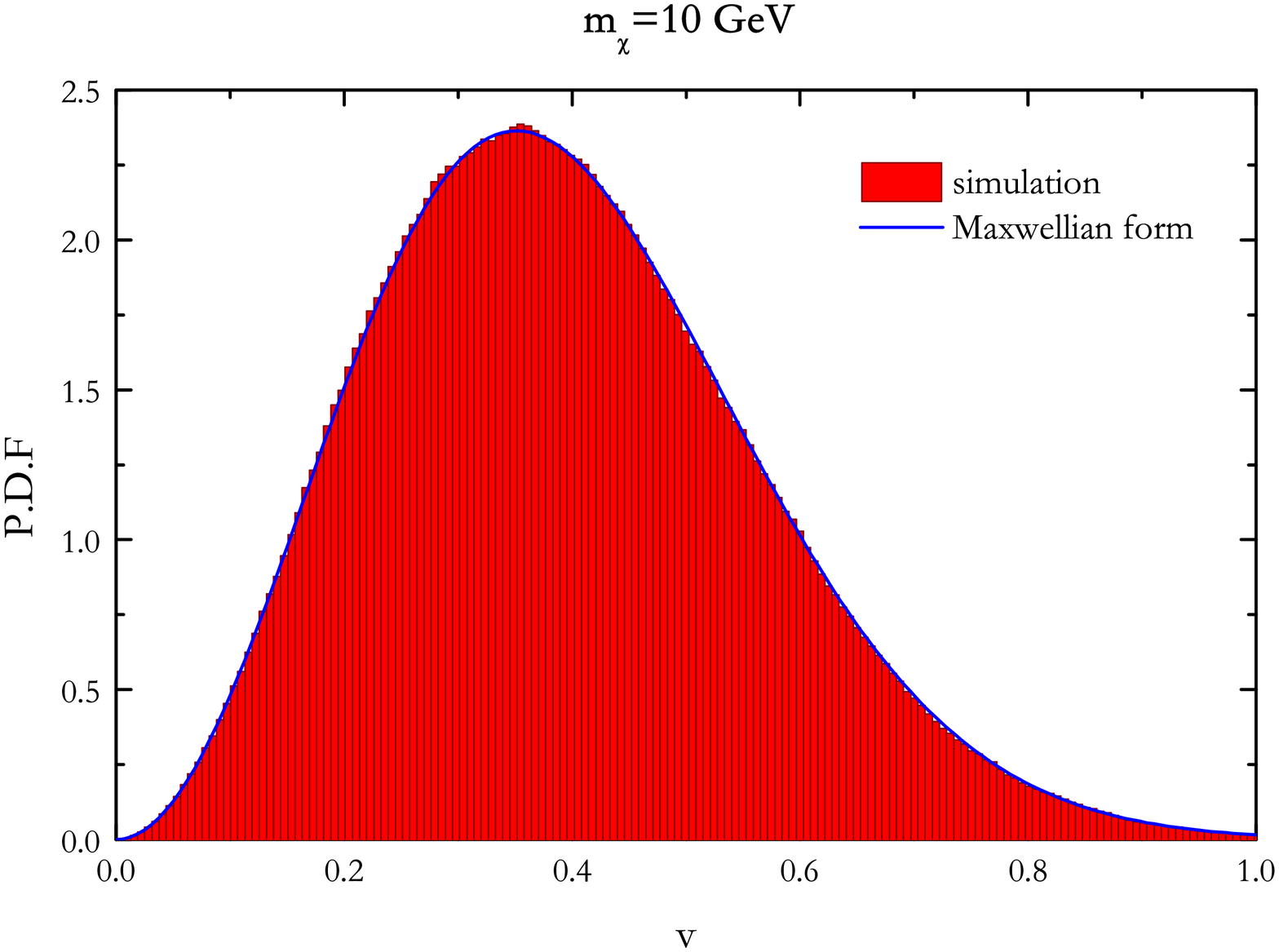}
\par\end{centering}

\protect\caption{\label{fig:appendixfig} \textbf{Left panel}: the normalized average kinetic energy $\left(2/3\right)\left\langle E_{\mathrm{k}}\right\rangle /T_{\chi}$
(solid) and the average square of the radial velocity component $\left\langle \left(v_{r}/v\right)^{2}\right\rangle $
(dotted) for DM mass $m_{\chi}=10\,\mathrm{GeV}$.  \textbf{Right panel}: The integrated  probability density function (P.D.F) of the solar DM particles obtained from the Brownian motion simulation (red histogram). For comparison also shown is the Maxwellian distribution (blue lines). See text for details.}
\end{figure}

To illustrate this point, we  first investigate the solar DM distribution
by following the ``Brownian motion'' method pioneered by the author of  Ref.~\citep{Nauenberg:1986em}.
A total number of $10^{5}$ successive collisions between the $10\,\mathrm{GeV}$
trial DM particle and solar elements are simulated. In the left panel of  Fig.~\ref{fig:appendixfig}
we present the quantity $\left(2/3\right)\left\langle E_{\mathrm{k}}\right\rangle /T_{\chi}$
and $\left\langle \left(v_{r}/v\right)^{2}\right\rangle $ from the
solar center up to $0.1\, R_{\mathrm{\odot}}$, a radius sufficient
to envelope almost all the DM particles trapped in the Sun (see Eq.~(\ref{eq:DMradius})). The curves  would be horizontal (at $1$ and $1/3$ respectively) for a thermal distribution.   We compare the simulated velocity distribution with the Maxwellian form in the right panel, where the velocity variable $v$ is nondimensionalized in terms of the velocity
unit $\left(GM_{\odot}/R_{\odot}\right)^{1/2}\approx436\,\mathrm{km\cdot s^{-1}}$.  It is evident   that the Maxwellian distribution
with the characteristic temperature $T_{\chi}$ indeed provides a good description
of the realistic  solar DM distribution for the  mass of
interest. $T_{\chi}=0.967\, T_{\mathrm{\odot}}(0) $ is determined from Eq.~(\ref{eq:EffectiveTemperature}). For larger DM masses, one can approximate $T_{\chi}$ as the  solar center temperature  $T_{\mathrm{\odot}}(0)$.

In addition, the Boltzmann form
is expected to remain an accurate description for the solar SIDM because
it naturally describes the thermal distribution of
the SIDM particles gathering in the solar core, while
the  effective temperature $T_{\chi}$ accounts for the non-thermal equilibrium
between the SIDM particles and the solar nuclei to a good approximation.
Therefore we still  adopt the Boltzmann distribution in the evaluation of self-capture and annihilation
in the Sun.

\section{\label{sec:appendixA}Self-evaporation rate of the trapped dark matter}

In this appendix we will verify the assumption made below Eq.~(\ref{eq:DMnumber equation})
that the evaporation rate due to the self-interaction of dark matter
is negligible compared to the annihilation rate. To this end we try
to estimate the dark matter self-evaporation rate in the approach
proposed in Ref.~\citep{Griest:1986yu}, by which we start with the
self-evaporation rate $\epsilon_{\mathrm{s}}$ in a unit volume :
\begin{equation}
\epsilon_{\mathrm{s}}=\frac{n_{\chi}^{2}}{2}\int d^{3}w_{1}\, f_{\chi}\left(\mathbf{w}_{1}\right)\int d^{3}w_{2}\, f_{\chi}\left(\mathbf{w}_{2}\right)\left|\mathbf{w}_{1}-\mathbf{w}_{2}\right|\int_{v_{\mathrm{esc}}}\frac{d\sigma\left(\mathbf{w}_{1},\mathbf{w}_{2};\mathbf{v}_{1},\mathbf{v}_{2}\right)}{dv_{1}dv_{2}}d^{3}v_{1}d^{3}v_{2},
\end{equation}
where $\mathbf{w}_{1}$ and $\mathbf{w}_{2}$ are respectively the
initial velocities of the trapped DM before scattering, with their
corresponding Maxwellian distributions $f_{\chi}\left(\mathbf{w}_{1}\right)$
and $f_{\chi}\left(\mathbf{w}_{2}\right)$ introduced in Eq.~(\ref{eq:velocity-distribution}),
$\mathbf{v}_{1}$ and $\mathbf{v}_{2}$ being the final velocity after
collision, one of which must exceed the local escape velocity $v_{\mathrm{esc}}$
in the Sun so as to be evaporated, and $d\sigma\left(\mathbf{w}_{1},\mathbf{w}_{2};\mathbf{v}_{1},\mathbf{v}_{2}\right)/dv_{1}dv_{2}$
schematically represents relevant differential cross section . Then
the conservation of energy and momentum implies
\begin{eqnarray}
f_{\chi}\left(\mathbf{w}_{1}\right)f_{\chi}\left(\mathbf{w}_{2}\right) & = & f_{\chi}\left(\mathbf{v}_{1}\right)f_{\chi}\left(\mathbf{v}_{2}\right),\nonumber \\
\left|\mathbf{w}_{1}-\mathbf{w}_{2}\right| & = & \left|\mathbf{v}_{1}-\mathbf{v}_{2}\right|,
\end{eqnarray}
so we have
\begin{eqnarray}
\epsilon_{\mathrm{s}} & = & \frac{n_{\chi}^{2}}{2}\int d^{3}v_{1}\, f_{\chi}\left(\mathbf{v}_{1}\right)\int d^{3}v_{2}\, f_{\chi}\left(\mathbf{v}_{2}\right)\left|\mathbf{v}_{1}-\mathbf{v}_{2}\right|\int\frac{d\sigma\left(\mathbf{v}_{1},\mathbf{v}_{2};\mathbf{w}_{1},\mathbf{w}_{2}\right)}{dw_{1}dw_{2}}d^{3}w_{1}d^{3}w_{2}\nonumber \\
 & \simeq & n_{\chi}^{2}\int_{v>v_{\mathrm{esc}}}d^{3}v\, f_{\chi}\left(\mathbf{v}\right)v\int\frac{d\sigma\left(\mathbf{v},0;\mathbf{w}_{1},\mathbf{w}_{2}\right)}{dw_{1}dw_{2}}d^{3}w_{1}d^{3}w_{2}\nonumber \\
 & = & n_{\chi}^{2}\int_{v>v_{\mathrm{esc}}}d^{3}v\, f_{\chi}\left(\mathbf{v}\right)v\sigma_{\mathrm{sc}}\left(v\right),\label{eq:cap-cs}
\end{eqnarray}
where we make approximation $\left|\mathbf{v}_{1}-\mathbf{v}_{2}\right|\simeq\left|\mathbf{v}_{1}\right|=v_{1}$
for the case $v_{1}>v_{\mathrm{esc}}\gg v_{2}$, in which the contribution
to evaporation rate doesn't suffer the Maxwellian tail suppression
of $f_{\chi}\left(\mathbf{v}_{2}\right)$, and $\sigma_{\mathrm{sc}}\left(v\right)$
is no other than the self-capture cross section with incident velocity
$v$. On the other hand, the annihilation rate in a unit volume is
written as
\begin{eqnarray}
n_{\chi}^{2}\left\langle \sigma_{\mathrm{ann}}v\right\rangle  & = & n_{\chi}^{2}\int d^{3}v_{1}\, f_{\chi}\left(\mathbf{v}_{1}\right)\int d^{3}v_{2}\, f_{\chi}\left(\mathbf{v}_{2}\right)\left|\mathbf{v}_{1}-\mathbf{v}_{2}\right|\sigma_{\mathrm{ann}}\left(\left|\mathbf{v}_{1}-\mathbf{v}_{2}\right|\right)\nonumber \\
 & \simeq & n_{\chi}^{2}\int d^{3}\left(v/\sqrt{2}\right)f_{\chi}\left(\mathbf{v}/\sqrt{2}\right)v\sigma_{\mathrm{ann}}\left(v\right)\nonumber \\
 & = & n_{\chi}^{2}\int d^{3}v\, f_{\chi}\left(\mathbf{v}\right)\sqrt{2}\,v\,\sigma_{\mathrm{ann}}\left(\sqrt{2}\,v\right),\label{eq:annilation-cs}
\end{eqnarray}
where $\sigma_{\mathrm{ann}}\left(v\right)$ is the annihilation cross
section with relative velocity $v$. Given that compared with Eq.~(\ref{eq:annilation-cs}) the integral over subspace $\int_{v>v_{\mathrm{esc}}}d^{3}v\, f_{\chi}\left(\mathbf{v}\right)$
in Eq.~(\ref{eq:cap-cs}) so overwhelmingly suppresses $\epsilon_{\mathrm{s}}$
in the DM mass region of our interest that we can safely ignore the
contribution from dark matter self-interaction in this study. Finally it is worth mentioning  that while our discussion is based on the assumption of a large mean free path for the SIDM particles, the self-evaporation will be heavily suppressed if the  optical depth is much smaller than the DM radius. Because in such case, the would-be evaporation events will end up being blocked by multiple collisions, which further favors our conclusion.

\providecommand{\href}[2]{#2}\begingroup\raggedright\endgroup

\end{document}